\documentclass{aa}  
\usepackage{graphicx}
\usepackage{txfonts}
\usepackage{amsmath2}

\newcommand{\pare}[1]{\left( #1 \right)}
\newcommand{\corc}[1]{\left[ #1 \right]}

\newcommand{\modu}[1]{\left | #1 \right |}

\newcommand{\eqn}[1]{\begin{equation} #1 \end{equation}}

\newcommand{\alt}{\lesssim}
\newcommand{\texto}[1]{{\mbox{\scriptsize{#1}}}}

\begin{document}
\title{Constraining Torus Models for AGNs Using X-ray Observations}
\subtitle{}
\author{E. Ibar
  \inst{1,2}
  \and
  P. Lira\inst{2}
}
\institute{
  Institute for Astronomy, University of Edinburgh,
  The Royal Observatory, Edinburgh, Scotland, EH9 3HJ.\\
  \and
  Departamento de Astronom\'ia, Universidad de Chile,
  Camino del Observatorio 1515, Santiago, Chile. \\
  \email{ibar@roe.ac.uk}
}
\date{
  Work in progress
}

\abstract
  {
    In Unification Models, Active Galactic Nuclei (AGN) are believed
    to be surrounded by an axisymmetric structure of dust and gas, which
    greatly influences their observed properties according to
    the direction from which they are observed. 
  }
  {
    The main aim of this work is to constrain the properties of this obscuring material
    using X-Ray observations. 
  }
  {
    The distribution of column densities observed by
    $Chandra$ in the {{\it{Chandra Deep Field South}}} is used to determine
    geometrical constraints for already proposed torus models.
  }
  {
    It is found that the best torus model is given by a classical
    `donut shape' with an exponential angular dependency of the
    density profile. The opening angle is strongly constrained by the
    observed column densities. Other proposed torus models are clearly
    rejected by the X-Ray observations.
  }
  {
  }
  
  \maketitle

\section{Introduction}
  
A unified scheme to explain a variety of Active Galactic Nuclei (AGNs)
has been proposed by Antonucci (\cite{antonucci}) to explain the
differences observed in Seyfert galaxies, a class of local active
galaxies. According to this {\it{Unification Model}}, the active
nucleus is surrounded by a toroidal structure composed of gas and
dust, which determines dramatic spectral differences depending on our
line of sight towards the central source. In the X-Ray domain, if an
AGN has a neutral Hydrogen column density ($N_H$) in the line of sight
smaller (when viewed face-on) or larger (when viewed edge-on) than
$10^{22}$ cm$^{-2}$, then the object is classified as Type I or Type
II, respectively.

The dusty toroidal structure that surrounds the central source absorbs
the optical, ultraviolet and soft X-Ray emission from the active
nucleus. However, hard X-Ray photons ($>2$ keV) can partially escape
from this cold material.  Therefore, from X-Ray observations it is
possible to measure the column density in the line of sight, and put
tight constraints on the properties of the obscuring medium.

It is commonly believed that this X-Ray emission is produced in a
'corona' near the super-massive black hole, where low energy photons
from the accretion disk are reprocessed by energetic electrons (either
mildly relativistic thermal electrons or highly relativistic
non-thermal electrons) via Inverse Compton Scattering. The main result
of this process is an observed spectral energy distribution described
by a power law, with a typical slope $\Gamma \approx 1.8$ (Turner et
al. \cite{turner}, Tozzi et al. \cite{tozzi}), and an exponential
cutoff energy at $\approx300$ keV (Matt et al. \cite{matt}). The
radiation can also be reflected and/or scattered depending on the
circumnuclear material distribution causing the overall spectral shape
to become flatter, $\Gamma\approx1.7$, with an apparent
smaller energy cutoff (see Svensson \cite{svensson}).

As hard X-Ray energy photons can penetrate the surrounding material,
and eventually escape from the AGN host, they can be detected with the
present generation of X-Ray telescopes. But if the Hydrogen column
density along the line of sight becomes larger than the inverse of the
photoelectric Cross Section $\sigma_{10{\rm keV}}^{-1}\approx 10^{24}$
cm$^{-2}$ (see Matt \cite{matt02}), then the medium becomes Compton
Thick ($\tau=N_H\sigma\approx1$). In this case, the only observable
emission component is coming from photons scattered and/or reflected
by the circumnuclear media (Wilman \& Fabian \cite{wil_fab}).

Since Compton Thin AGNs (those with $N_H<10^{24}$ cm$^{-2}$) are
ubiquitous emitters in the $2-10$ keV hard X-Ray band, deep X-Ray
surveys have proven to be the best way to estimate the number density
and evolution of active galaxies in the Universe (Mushotzky
\cite{Mushotzky}). Unfortunately, these X-Ray surveys are always flux
limited (nowadays $F^\texto{limit}_\texto{2-10keV}\approx10^{-16}$ erg
sec$^{-1}$ cm$^{-2}$), and therefore biased against faint sources. In
fact, the population of high redshift, low luminosity, and highly
obscured AGNs is still not well determined. This problem translates
into a poorly determined faint end of the luminosity function,
implying uncertainties in the contribution of obscured AGNs to the
Cosmic X-Ray Background (CXRB), which is known to require of a sizable
population of obscured sources in order to explain its hard spectral
energy distribution.

The aim of the present work is to constrain different torus geometries
that can reproduce the observed distribution of Hydrogen column
densities ($N_H$) in AGNs observed by $Chandra$ in the {\it{Chandra
Deep Field South}} (CDF-S).

Section 2 describes the intrinsic $N_H$ distribution found by Tozzi et
al. (\cite{tozzi}) which was used in our analysis. In Section 3 we
explain the theoretical modelling, while in Section 4 we present our
models and results. In Section 5 our results are discussed and we
present the conclusions in Section 6. Throughout this work we use:
$\Omega_m=0.3$, $\Omega_\Lambda=0.7$ and $H_0=70$ km sec$^{-1}$
Mpc$^{-1}$ (Spergel et al. \cite{spergel}).

\section{Deep X-Ray Observations}
\subsection{AGNs in the $Chandra$ Deep Field-South}

In the past few years, several deep $Chandra$ images of the
extragalactic sky have been obtained, with the $2\ Ms$ $Chandra$ Deep
Field-North (CDF-N; Alexander et al. \cite{Alex}) and $1\ Ms$
$Chandra$ Deep Field-South (CDF-S; Giacconi et al. \cite{Giac}) being
the two deepest ($F^\text{limit}_{2-10\ keV}\approx 1.4\cdot 10^{-16}$
erg s cm$^{2}$ for CDF-N and a factor 2 shallower for the
CDF-S). Optical identifications have been obtained for most X-Ray
sources in these fields. Barger et al. (\cite{Barger03}) presented
multicolour imaging data for all 503 X-Ray point sources in the CDF-N
and spectroscopic redshifts for $\sim 56\%$ of the sample. Zheng et al
(\cite{zheng}) presented accurate photometric redshifts for 342
sources in the CDF-S using the large multicolour data set available
for this field. This corresponds to $\sim 99\%$ of the sample and
includes 173 sources with reliable spectroscopic redshifts from
Szokoly et. al (\cite{Szok}).

Given the photo-z completeness of $\sim 99\%$ for the the X-Ray sources
in the CDF-S, this field becomes the perfect choice for our study of
the torus properties. Redshifts are necessary in order to compute the
observed $N_H$ (see below) as well as the source X-Ray luminosity.

Following the sample definition for the CDF-S found in Tozzi et al.
(\cite{tozzi}), out of the 347 X-Ray detected sources (346 sources
presented by Giacconi et al. (\cite{Giac}) and 1 extra source
presented by Szokoly et al. (\cite{Szok})), 7 sources are stars and 4
have no redshift information. Further 15 sources have X-Ray
luminosities $< 10^{41}$ erg s$^{-1}$ cm$^{-2}$, a luminosity range
where emission from star forming galaxies becomes significant, and are
therefore discarded. Finally, 14 Compton Thick AGN were also
eliminated, as discussed in Section 4. The final sample contains 307
X-Ray point sources at $z>0$ that will be used for our study.

\subsection{Hydrogen Column Densities}

The distribution of $N_H$ values for the sources in the CDF-S were
derived by Tozzi et. al (\cite{tozzi}). They performed direct X-Ray
spectral fitting to the observed Chandra data using the redshift
information published by Zheng et. al (\cite{zheng}). Using a
subsample with the 82 brightest sources, Tozzi et. al (\cite{tozzi})
determined a weighted mean spectral index of $\Gamma \sim 1.75$ which
resulted independent of the values of the derived $N_H$. By fixing
$\Gamma = 1.8$ the redshifted $N_H$ was derived for the rest of the
sample. The distribution was then corrected for incompleteness in the
luminosity and redshift parameter space using the survey effective
area and the luminosity function of Ueda et al. (\cite{Ueda}). Ueda's
luminosity function was determined using 247 AGN detected in the
$2-10$ keV X-Ray band from a compilation of $ASCA$, $HEAO-1$ and
$Chandra$ observations. The luminosity function shows a luminosity
dependent density evolution, in which the low luminosity AGN
population peaks at a lower redshift than the high luminosity
sources. This is consistent with optical quasar observations in which
this population peaks at redshift $\sim2$ (Boyle et al. \cite{boyle}).
 
The final corrected $N_H$ distribution, seen in Figure 16 of Tozzi
et. al (\cite{tozzi}), will be used in the following sections to
constrain the torus parameters.

\section{Model General Properties
\label{teo_back}}

We first assume that the X-Ray emission in AGNs comes from the nearest
region to the super-massive black hole. This region has a typical
spatial scale of $R_C\approx 10-100\ R_S$ (where $R_S=GM_{BH}/c^2$ is
the Schwarzschild Radius), implying physical sizes of $R_C\approx
10^{-5}-10^{-3}$ pc for a typical super-massive black hole with
$M_{BH}\approx10^7-10^{8}\ M_\odot$. Recent observations of NGC\,1068
using mid-infrared interferometry, suggest a torus about $3.4$ pc
diameter (Jaffe et al. \cite{jaffe}), while near-IR reverberation
mapping of nearby Seyfert galaxies estimates an inner limit at $\sim
10^{-2}-10^{-1}$ pc (Suganuma et al. \cite{Suganuma06}). Therefore,
for practical purposes we can approximate the X-Ray source as a
point-like emitting region, implying an easy treatment for the X-Ray
radiation that will provide the total column density along the line of
sight.

\subsection{Torus Properties
\label{section3}}

We consider four different torus models (presented in Section
\ref{torus_models}) to obtain synthetic distributions of column
densities which are then compared with the observed distribution from
Tozzi et. al (\cite{tozzi}). 

We model the obscuring region as a simple axisymmetrical matter
distribution surrounding the accretion disk. Given the symmetry of the
problem, random lines of sight with polar angles between 0 and $\pi/2$
(weighted by the angle differential area), are used to calculate the
optical depth for each random direction, where the the polar angle
$\phi$ is defined as the angle subtended between the line of sight and
the torus equatorial plane.

We assume a solar abundance for the torus, a photoelectric cross
section for absorption given by Morrison \& McCammon (\cite{MoMc}), a
standard Galactic {\it{Gas-to-Dust}} ratio
$N_H/E(B-V)=5.8\cdot10^{21}$ cm$^{-2}$ given by Bohlin et
al. (\cite{Bohlin}) and a solar neighbourhood value $R_V=3.1$ (Schultz
\& Weimer \cite{sch_wei}). These assumptions relate the Hydrogen
column density $N_H$ with the optical extinction $A_V$, and the
optical depth $\tau_V$ ($A_V=1.09 \tau_V$),

\eqn{
  \label{nhss}
  \begin{split}
    N_H(\phi) &= \sigma^{-1} \tau =\int \rho(r,\phi)  dr  \\ 
    &=5.8\cdot10^{21}\ E(B-V)\ {\rm cm}^{-2}= 1.9\cdot 10^{21}A_V\ {\rm cm}^{-2}\\  
    &=2.0\cdot 10^{21}\tau_{V}\ {\rm cm}^{-2}= 2.0\cdot 10^{21}\ \tau_{Vi}\
    N_T\ {\rm cm}^{-2} 
  \end{split}
}

\noindent
where, if the medium is continuous, $\rho(r,\phi)$ is the matter
density distribution and $\sigma$ the effective absorption cross
section. Otherwise, if the medium is clumpy, $\tau_{Vi}$ represents
the optical depth for one cloud and $N_T$ the average number of clouds
along the line of sight. It is easy to show that if the density
distribution $\rho$ has no angular dependency (i.e., $\rho=\rho(r)$),
then the radial dependency will be hidden by the assumed maximum
column density $N_{Hmax}$ along the line of sight (see Section
\ref{torus_models}).  

Other model parameters, such as the inner radius, external radius,
density in the equatorial plane, and/or cloud optical depth, are also
constrained by $N_{Hmax}$. In this work we fix the maximum value for
the column density at $N_{Hmax}=10^{25}$ cm$^{-2}$, according to X-Ray
observations of Type II AGNs in the $0.1-100$ keV band using the
BeppoSAX X-Ray satellite (Maiolino et al. \cite{Maiolino}).

Due to the degeneracy of the model parameters, we use the minimum
number possible of free parameters to describe the torus
distribution. The chosen parameterisations do not give physical
sizes of the distributions explicitly, but instead reduced parameters
such as $R_{int}/R_{out}$ (where $R_{out}$ is the external radius and
$R_{int}$ is the inner radius for the torus) are used. $R_{int}$ can
be estimated by the `evaporation radius' for graphites (at $T\approx
1500\ K$) defined by assuming thermal equilibrium between the
ultraviolet incident radiation and the rate for reemission from the
dust. Assuming $a_\texto{gr}=0.05\ \mu$m as a typical grain size, and
following Barvainis (\cite{bar}),

\eqn{
  \label{radio_min}
  R_{int}\approx0.4 L_{UV,45}^{1/2}\  \texto{\small{pc}}
,}

\noindent
where $L_{UV,45}$ is the ultraviolet (UV) luminosity per unit of
$10^{45}$ erg sec$^{-1}$. Assuming this value as a lower limit for the
distance, we can estimate the external radius $R_{out}$, and other
model parameters using our torus model results (see estimations in
Table \ref{table1}).

\section{Modelling the Torus 
\label{torus_models}}

The large degeneracy of model parameters and the uncertainties given
by the lack of faint sources in the $N_H$ distribution determined by
Tozzi et al. (\cite{tozzi}), require us to keep the number of free
parameters within our torus models to a minimum. As such, we do not
allow the parameters to change with luminosity or redshift.

Barger et al. (\cite{Barger05}) presented a complete sample of local
X-Ray selected AGN ($z<1.2$), in which a luminosity dependence for the
fraction of Type II over Type I AGNs is seen, as Ueda et
al. (\cite{Ueda}) proposed earlier. This implies a possible dependency
of the covering factor of the dusty torus on luminosity, and therefore
a change in the geometrical torus distribution. Hence, our results can
be understood as an average geometrical distribution for the entire
AGN population, given by a simple axially symmetric geometry, with no
dependency on luminosity or redshift. A discussion on the luminosity
dependency can be found in Section \ref{discussion}.

To determine the torus parameters which best fit the observed column
densities, we use the $\chi^2$ test assuming poissonian error bars for
the observed data. Our own experience determining the intrinsic $N_H$
distribution based on {\it Hardness Ratio} measurements suggests that
this method introduces large uncertainties in the inferred $N_H$
values, which dominate over the poissonian errors from a population of
sources. On the other hand, the spectral treatment introduced by Tozzi
et al. (\cite{tozzi}) gives more reliable values for the intrinsic
properties of the obscuring region present in AGNs.

Tozzi et al. (\cite{tozzi}) found 14 sources showing a
reflection-dominated spectrum. These Compton Thick sources were not
used in our analysis because of the large uncertainties in the number
counts of this population. In fact, we have not used the last bin in
the intrinsic column density distribution shown in Figure 16 of Tozzi
et al. \cite{tozzi}, which includes objects with $N_H > 10^{24}$
cm$^{-2}$ and corresponds to a lower limit only.

The models presented in the remaining of this section were motivated
by previous work which were carried out to explain a variety of
observational properties of AGNs (Treister et al. \cite{Treister};
Pier \& Krolik \cite{pi_kr92}, \cite{pi_kr93}; Granato \& Danese
\cite{gr_da}; Elitzur et al. \cite{Elitzur}). A summary of the results
from the modelling is presented in Table \ref{table1}.

\subsection{Model 1}

Following the torus distribution proposed by Treister et
al. (\cite{Treister}), we modelled the geometry given by Figure
\ref{torus}a to find different optical depths as a function of the
line of sight, $\phi$. In this case the matter density is not
distributed as in an homogeneous medium, but has the following
dependence with $\phi$ and $r$:

\eqn{
  \rho(r,\phi) = \rho_\texto{eq} \pare{\frac{r_\texto{int}}{r}}^\beta\
  e^{-\gamma\modu{\sin(\phi)}}
}

\noindent
where $\rho_\texto{eq}$ is the inner density (at $r_\texto{int} =
R_m-R_t$) in the equatorial plane (constrained by $N_{Hmax}$, see
Equation \ref{nhmax1}). In this model, the free parameters are
$R_m/R_t$ (see Figure \ref{torus}a), $\gamma$, and $\beta$. The
analytical equations which describe the column density as a function
of $\phi$ are given by: 

\begin{flushleft}
\begin{equation}
\label{equ1}
\begin{array}{rl}
&\texto{\small{If}} \quad 0\leq\phi  < \sin^{-1}\pare{R_t/ R_m} \quad
  \texto{\small{and}} \quad \beta\neq1 \quad \texto{\small{then,}}\\ 
&N_H = N_{Hmax}\ e^{-\gamma\modu{\sin(\phi)}}\ \times \\
&\quad\quad
\frac{\corc{\pare{\cos(\phi)+\sqrt{(R_t/R_m)^2-\sin(\phi)^2}}^{1-\beta}-
  \pare{\cos(\phi)-\sqrt{(R_t/R_m)^2-\sin(\phi)^2}}^{1-\beta}}} 
  {(1+R_t/R_m)^{1-\beta}-(1-R_t/R_m)^{1-\beta}}\\ 
&\texto{\small{If}} \quad 0\leq\phi  < \sin^{-1}\pare{R_t/R_m} \quad 
\texto{\small{and}} \quad \beta=1 \quad \texto{\small{then,}}\\
&N_H = N_{Hmax}\ e^{-\gamma\modu{\sin(\phi)}}\ \times \\
&\quad\quad
\frac{\ln\pare{\pare{\cos(\phi)+\sqrt{(R_t/R_m)^2-\sin(\phi)^2}}/
\pare{\cos(\phi)-\sqrt{(R_t/R_m)^2-\sin(\phi)^2}}}}
     {\ln\pare{(1+R_t/R_m)/(1-R_t/R_m)}}\\
&\texto{\small{If}} \quad \sin^{-1}\pare{R_t/R_m} \leq \phi \leq
  \pi/2\quad 
\texto{\small{then,}} \quad N_H = N_{Hmin}
\end{array}
\end{equation}
\end{flushleft}

The maximum and minimum column density along the line of sight is
through the equatorial plane and at the poles, respectively.

\eqn{
  \begin{split}
    \label{nhmax1}
    \texto{\small{If}}\quad &\beta\neq1 \quad \texto{\small{then,}} \quad\\
    &N_{Hmax} =  \rho_\texto{eq}\ r_\texto{int}^\beta
    \frac{R_m^{1-\beta}}{1-\beta}  
    \corc{\pare{1+\frac{R_t}{R_m}}^{1-\beta}-
      \pare{1-\frac{R_t}{R_m}}^{1-\beta}}\\  
    \texto{\small{If}}\quad &\beta=1\quad\texto{\small{then,}} \quad
    N_{Hmax} =  \rho_\texto{eq}\ r_\texto{int}
    \ln\pare{\frac{1+R_t/R_m}{1-R_t/R_m}}\\
  \end{split}
}

This model gives a good fit to the data, with $\chi^2_{\nu,min}=1.30$.
Results adopting $\beta=0.0$ are presented in Figure \ref{ks_models}
and show a well constrained region of the $R_m/R_t - \gamma$ parameter
space. The best fit values are spread between $1.07\alt R_m/R_t \alt
1.15$ and $7.0 \alt \gamma \alt 9.0$ at 3 $\sigma$ confidence
levels. The same results are obtained when a value $\beta=2.0$ is
adopted. Adopting $\beta=1.0$ and $3.0$ results in a small decrease in
the value of $\gamma$, but does not introduce further changes. Note
however, that $\beta$ modifies the value of $N_{Hmax}$ through
Equation \ref{nhmax1}.

From the family of parameters found, we show in figure
\ref{best_models} some examples of synthetic $N_H$ distributions and
compared them with the one determined by Tozzi et al. (\cite{tozzi}).
It is seen that the parameter $R_m/R_t$ mostly affects the
distribution of Type I sources (i.e., for $N_H < 10^{22}$ cm$^{-2}$),
while $\gamma$ changes the Type II/I fraction.

A sketch of the torus geometrical distribution, choosing the best fit
parameters, is given in Figure \ref{geo_models}, where the `donut'
structure of the model is clearly seen. Under these assumed
parameters, this model predicts an intrinsic fraction of sources with
column densities $>10^{24}$ cm$^{-2}$ of $\sim27\%$.

\begin{table*}
\begin{center}
\begin{tabular}{|c|c|c|c|c|c|c|}
  \hline
  Model & 
  $\chi^2_{\nu,min}$ & 
  Parameter 1 & 
  Parameter 2 & 
  Parameter 3 & 
  $R_\texto{out}[pc]$ & 
  $\rho_\texto{eq}[cm^{-3}]$ \\ 
  \hline
  1 & 
  $1.30$ &
  $1.07\alt R_m/R_t\alt 1.15$ &
  $7.0\alt \gamma\alt 9.0$& 
  $0.0\alt \beta \alt3.0\phantom{.}^\ddag$ & 
  $5.7 - 12$ & 
  $2.8 - 6.0 \cdot10^{5}\phantom{.}^\dagger$\\
  \hline
  2 &  
  $8.13$ &
  $2.4\alt h/a\alt 3.3$ &
  $250\alt b/a \alt 600$ & 
  X & 
  $100 - 240$ & 
  $1.3 - 3.2 \cdot10^{4}$\\
  \hline
  3 &
  $5.65$ & 
  $1.0\alt\phi_{c}\ rad \alt1.2$ &
  $8.0\alt \gamma \alt 10.0$ & 
  $free$ & 
  $1.0\phantom{.}^\ast$ &
  $5.4\cdot10^6 \phantom{.}^\dagger$\\
  \hline
  4 &  
  $2.65$ &
  $0.38\alt \sigma\ rad \alt 0.44$ & 
  $\tau_{Vi} \alt 0.8$ & 
  X & 
  $free$& 
  -\\
  \hline
\end{tabular}
\end{center}
Notes .-\\ 
$\phantom{.}^\ast$ indicate a fixed parameter.\\
$\phantom{.}^\dagger$ these parameters can be modified taking into
account radial dependencies for the model, i.e. $\beta\neq0$. \\
$\phantom{.}^\ddag$ obtained assuming $\gamma=9.0$.
\caption
{ Summary of the estimated parameters for the models described in
  Section \ref{torus_models}. Ranges are obtained from 3 $\sigma$
  confidence limits (see contour plots in the Figure \ref{ks_models}).
  The last two columns are deduced considering a UV luminosity of
  $10^{45}$ erg sec$^{-1}$ and Equation \ref{radio_min}. No radial
  dependency for the density distribution ($\beta=0.0$) is assumed in
  the models. In Model 3 we use a fixed parameter $R_\texto{out}=1$ pc
  to estimate a density at the equatorial plane, although this is a
  free parameter. Model 4 is not shown in the last columns due to the
  different cloud treatment explained in Subsection \ref{model4}.  }
\label{table1}
\end{table*}

\subsection{Model 2}

This model is proposed following previous studies by Pier \& Krolik
(\cite{pi_kr92}, \cite{pi_kr93}) based on the IR reemission coming
from the torus. We use their geometrical distribution (Figure
\ref{torus}b) to find the dependency of the column density with the
line of sight.

This model assumes a homogeneous density with $h/a$ and $b/a$ as free
parameters, where $h$, $b$, and $a$ are as shown in Figure
\ref{torus}b. The dependency of the column density with the line of
sight ($0\leq\phi\leq\pi/2$) is given by:

\begin{center}
\begin{equation}
\begin{array}{rl}
\texto{\small{If}}& \quad 0\leq\phi<\tan^{-1}\pare{\frac{h}{2b}} \quad
\texto{\small{then,}}\quad
N_H=N_{Hmax}\frac{\cos\pare{\tan^{-1}\pare{h/2b}}}{\cos(\phi)}\\ 
\texto{\small{If}}& \quad
\tan^{-1}\pare{\frac{h}{2b}}\leq\phi<\tan^{-1}\pare{\frac{h}{2a}}
\quad \texto{\small{then,}}\quad\\ 
&N_H=N_{Hmax}\frac{\cos\pare{\tan^{-1}\pare{h/2b}}}{b/a-1}
\pare{\frac{h}{2a}\sin(\phi)^{-1}-\cos(\phi)^{-1}}\\   
\texto{\small{If}}& \quad \tan^{-1}\pare{\frac{h}{2a}}\leq\phi\leq
\pi/2 \quad \texto{\small{then,}}\quad 
N_H=N_{Hmin}
\end{array}
\end{equation}
\end{center}

\noindent
where the maximum column density is:

\eqn{
  N_{Hmax}=\rho_\texto{eq}\
  \frac{b-a}{\cos\pare{\tan^{-1}\pare{\frac{h}{2b}}}} 
}

Comparing the synthetic distributions of column densities with the
observed one, we found the results shown in Figure
\ref{ks_models}. The parameters are not well constrained and imply
$2.4\alt h/a \alt 3.3$ and $250\alt b/a \alt 600$ at 3 $\sigma$
confidence levels. The model also shows an extreme radial-to-vertical
thickness ratio ($b/h\approx 75-250$).

Synthetic $N_H$ histograms are shown in Figure \ref{best_models}. It
is seen that the parameter $b/a$ systematically overestimates the
number of sources at around $N_H=10^{22.5}\ cm^{-2}$, while $h/a$
modifies the first bin of unobscured sources.

Clearly, this model is limited by its constant density distribution.
This translates into a large $b/a$, needed to reproduce the wide range
of observed column densities (since $N_H\propto \Delta s$), and a very
sensitive distribution in the innermost region of the torus (see
Figure \ref{geo_models}). Also, the model does not predict a large
fraction of highly obscured sources, implying that it cannot explain
the CXRB or the number of Compton Thick sources observed in the CDF-S
($\sim 10\%$; Tozzi et al.  \cite{tozzi}). The typical intrinsic
fraction of Compton Thick sources predicted by the model is only
$\sim3\%$ which are produced by a narrow angular section near the
equatorial plane able to account for large column densities (see
Figure \ref{geo_models}).

When we compare our analysis with the results by Pier \& Krolik
(\cite{pi_kr92}), we find further disagreements. Typical dimensions
deduced by them, as restricted by IR torus reemission, are of
parsec-scale, meanwhile we obtain a torus with typical scales of
$\sim200$ pc assuming $L_{UV}=10^{45}$ erg sec$^{-1}$.

We conclude that this torus model can be rejected due to the numerous
caveats already discussed, such as typically large outer radius, which
is not seen in local AGNs (Jaffe et al. \cite{jaffe}; Prieto et al
\cite{prieto}), the sensibility of the geometrical distribution in the
inner regions, and the small number of predicted Compton Thick sources.

\subsection{Model 3}

This model has the spatial density distribution shown in Figure
\ref{torus}c and is based on the work presented by Granato \& Danese
(\cite{gr_da}). The density profile has an exponential dependency with
the polar angle parameterised by a factor $\gamma$ (when
$\phi<\phi_c$), and a power law dependancy with the radius
parameterised by the index $\beta$:

\eqn{
  \rho(r, \phi) = \rho_\texto{eq} \ 
  \pare{\frac{r_\texto{int}}{r}}^\beta \ e^{-\gamma\modu{\sin(\phi)}} 
}

\noindent
where $\rho_\texto{eq}$ is the inner torus density at the equatorial
plane. The free parameters of the model are the opening angle
$\phi_{\texto{c}}$, $\gamma$ and $\beta$. The equation which describes
the dependence of the column density with the line of sight is:

\eqn{
\begin{split}
\texto{\small{If}}& \quad 0 \leq \phi < \phi_\texto{c} \quad
\texto{\small{then,}}\quad N_H=N_{Hmax} \
e^{-\gamma\modu{\sin(\phi)}}\\ 
\texto{\small{If}}& \quad \phi_\texto{c} \leq \phi \leq \pi/2 \quad
\texto{\small{then,}}\quad N_H=N_{Hmin}\\
\end{split}
}

\noindent
where $N_{Hmin}$ and $N_{Hmax}$ are fixed parameters at $10^{20}$ and
$10^{25}$ cm$^{-2}$, respectively. The minimum value fixes $N_H$ at the
poles and the maximum value fixes $N_H$ at the equatorial plane. 

\eqn{
\begin{split}
\label{nhmax3}
  \texto{\small{If}}& \quad \beta\neq1\quad \texto{\small{then,}}\quad
  N_{Hmax} =\rho_\texto{eq}\ r_\texto{int}^{\beta} 
  \pare{\frac{r_\texto{out}^{1-\beta}-r_\texto{int}^{1-\beta}}{1-\beta}}\\ 
  \texto{\small{If}}& \quad\beta=1\quad \texto{\small{then,}}\quad
  N_{Hmax} = \rho_\texto{eq}\
  r_\texto{int}\ln\pare{\frac{r_\texto{out}}{r_\texto{int}}} 
\end{split}
}

Searching for best fit parameters using the $\chi^2$ test, we obtain
the results shown in Figure \ref{ks_models}. This model gives a poor
fit to the data ($\chi^2_{\nu,min}=5.65$), and is not able to
reproduce the observed $N_H$ distribution between the first bin and
those at $\la 10^{22}$ cm$^{-2}$ (Figure \ref{best_models}). The sharp
gap seen at $N_H \alt 10^{21}cm^{-2}$ in the synthetic distributions
corresponds to the transition produced when the line of sight no
longer hits the torus ($\phi >\phi_\texto{c}$). This dichotomy also
affects the $\chi^2$ distribution where multiple minima are
seen. Parameters between $57^\texto{o} \alt \phi_\texto{c} \alt
69^\texto{o}$ and $8.0\alt \gamma \alt 10.0$ are found.

Simulations of the intrinsic $N_H$ distributions are shown in Figure
\ref{best_models}. Changes in the value of $\phi_\texto{c}$ affect the
distribution over Type I sources only (i.e., for $N_H < 10^{22}$
cm$^{-2}$), while $\gamma$ affects the number of sources at $N_H
\approx 10^{21}$ cm$^{-2}$ and changes the fraction of Type II/I. The
estimated intrinsic fraction of Compton Thick sources, using the best
fit parameters, is about $26\%$.

When we compare our results with the work by Granato \& Danese
(\cite{gr_da}), we find that our analysis can constrain $\phi_c$
(except for large $\gamma$ values), while Granato \& Danese
(\cite{gr_da}) found that the spectral energy distribution for the
reemission does not depend strongly on this parameter. On the other
hand, we cannot constrain the outer radius, since this scale parameter
is contained in $N_{Hmax}$ (Equation \ref{nhmax3}), and it could be
easily adjusted to adopt any value (e.g. $>100$ pc, as found by
Granato \& Danese \cite{gr_da}).

\subsection{Model 4
\label{model4}}

Many previous theoretical models trying to reproduce the IR reemission
coming from the obscuring torus were based on continuous density
distributions (as those presented in Models 1 to 3). Elitzur et
al. (\cite{Elitzur}) and Nenkova et al. (\cite{Nenkova}) have
developed the formalism to handle dust clumpiness and their results
indicate that the inclusion of thick clouds may resolve the
difficulties encountered by previous theoretical efforts trying to
reproduce the observed IR spectral energy distributions of local
AGNs. They have shown that some problems are naturally resolved if a
clumpy media with $\sim5-10$ clouds along radial rays (each one with
$\tau_{V_i}\gtrsim40$) is considered. The fundamental difference
between clumpy and continuous density distributions is that radiation
can propagate freely between different regions of an optically thick
medium when it is clumpy, but not otherwise.

Following Elitzur et al. (\cite{Elitzur}) we define the model given by
Figure \ref{torus}d, considering a spatially random cloud distribution
per unit length given by:

\eqn{
  n_T(r,\phi)=n_{\texto{eq}}\pare{\frac{r_\texto{int}}{r}}^\beta\
  e^{-\phi^2/\sigma^2} 
}

\noindent
where $n_{\texto{eq}}$ is the inner number of clouds by unit length in
the equatorial plane, $\sigma$ the half opening angle for the angular
distribution, and $\beta$ the power law index describing the radial
dependency. Integrating the last equation along the line of sight, we
obtain the average number of clouds, $N_T$ as a function of the line
of sight, parameterised by the angle $\phi$:

\eqn{
  N_T(\phi)=N_{\texto{eq}}\ e^{-\phi^2/\sigma^2}
}

\noindent
where $N_\texto{eq}$ is the average number of clouds along the
equatorial plane: 

\eqn{
\begin{split}
\label{nhmax5}
\texto{\small{If}}& \quad \beta\neq1 \quad \texto{\small{then,}}\quad
N_{\texto{eq}}= n_{\texto{eq}}\frac{r_\texto{int}^\beta}{1-\beta}
\pare{r_\texto{out}^{1-\beta}-r_\texto{int}^{1-\beta}}\\
\texto{\small{If}}& \quad \beta=1 \quad \texto{\small{then,}}\quad
N_{\texto{eq}}= n_{\texto{eq}}
r_\texto{int}\ln\pare{\frac{r_\texto{out}}{r_\texto{int}}}\\ 
\end{split}
}

Assuming $\tau_V = N_T\tau_{V_i}$, and considering Equation
\ref{nhss}, we can estimate the Hydrogen column density dependency on
the line of sight, as:

\begin{equation}
  N_H(\phi)=N_{Hmax} \ e^{-\phi^2/\sigma^2}
\end{equation}

\noindent
The fixed parameter $N_{Hmax}$ constrains the optical depth, the
number of clouds along the line of sight at the equatorial plane, and the
parameter $\beta$: 

\eqn{
  \label{nhmax4}
  N_{Hmax}=2.0\cdot10^{21} N_\texto{eq}\tau_{Vi}
}

If a certain value $\tau_{Vi}$ is assumed for the clumps, the only
remaining free parameter for this distribution is $\sigma$. However,
restricting $N_T$ to represent a discrete number of clouds and
assuming a poissonian probability for the number of clouds along the
line of sight, it is also possible to constrain the optical depth per
cloud, $\tau_{Vi}$. In the case when the optical depth tends to zero
but $N_T$ tend to infinity, the continuous case is recovered, where
column densities are allowed to have any value. Note that Nenkova et
al. (\cite{Nenkova}) assumed $\tau_{V_i}=40$ to reproduce the IR
reemission from the torus, implying a minimum column density of
$8.0\cdot10^{22}$ cm$^{-2}$. Clearly, smaller optical depths per cloud
are necessary to explain the observed distribution of $N_H$.  A work
around this difficulty which would still be consistent with a large
$\tau_{Vi}$ per cloud, is to assume that absorption by overdensities of
diffuse material located in the outer regions or within the torus
might be responsible for the smaller observed column densities.

The best-fit values for $\sigma$ and $\tau_{Vi}$ are shown in Figure
\ref{ks_models}. Multiple minima in the $\chi^2$ distribution are
found due to the cloud number discretization and the selected $N_H$
bin-size. Values between $22^\texto{o} \alt \sigma \alt 25^\texto{o}$
and $\tau_{Vi}\alt 0.8$ are preferred. Parameter $\sigma$ changes the
number of objects at $N_H=10^{20}$ cm$^{-2}$ and at $N_H=10^{25}$
cm$^{-2}$, without altering significantly the intermediate values
where a increasing slope in the synthetic distributions is seen. On
the other hand, the optical depth per cloud creates a dichotomy
between Type II and Type I objects, where the minimum value of $N_H$
is given by the optical depth of a single cloud.

From Figure \ref{geo_models}, it is clear that this model gives the
largest number of Compton Thick sources produced by lines of sights
near the equatorial plane (approximately a fraction of $\sim 58\%$
using best fit parameters). On the other hand, if a lower $N_{Hmax}$
of $10^{24}$ cm$^{-2}$ is adopted, the gaussian angular dependency
generates large discrepancies with the observed column densities.

When we compare our parameters with the results coming from the
reemission treatment given by Elitzur et al. (\cite{Elitzur}), we
found that our typical Gaussian angular distributions are smaller than
Elitzur's values. In fact, they find a valid range of
$\sigma=45^\texto{o}\pm15^\texto{o}$. Also, their estimations for the
number of clouds along the equatorial plane ($N_\texto{eq}\approx
5-10$) correspond to a maximum column density of $\sim10^{24}$
cm$^{-2}$ (see Equation \ref{nhss}), one order of magnitude smaller
than our assumptions. In our models, since $N_{Hmax}=10^{25}$
cm$^{-2}$, the value of $N_\texto{eq}$ depends on the assumed
$\tau_{Vi}$.

Note that since $N_H\propto N_T$, this clumpy model requires a wide
range for the number of clumps (at least $\sim3$ orders of magnitude)
to describe the wide range of observed $N_H$s.

\section{Discussion
\label{discussion}}

\subsection{Torus Properties as a Function of Luminosity}

The work of Barger et al. (\cite{Barger05}), based on a complete AGN
sample at $z<1.2$, shows a dependency of the fraction of Type II/I AGN
with luminosity. This observational evidence implies that our simple
models, where the torus does not evolve with neither luminosity nor
redshift, might not be an accurate representation.

Treister \& Urry (\cite{Treister05}) found a simple explanation for
Barger's results assuming a {\it{Modified Unification Model}}, where
the percentage of obscured AGNs varies linearly from 100\% at
$L_X=10^{42}$ erg sec$^{-1}$ to 0\% at $L_X=3\cdot10^{46}$ erg
sec$^{-1}$. We can examine how this luminosity evolution would change
our results for the torus models. Using the luminosity dependency for
the fraction of Type II AGN determined by La Franca et
al. (\cite{lafranca05}), we find that in the case of Model 1 (the one
with the best statistical results), parameter $\gamma$, which is
mostly responsible for the variations observed in this fraction,
changes to $\approx7.0$, $14$ and $23$ (fixing $R_m/R_t=1.1$) for the
luminosity bins $10^{42-43}$, $10^{43-44}$ and $10^{44-45}$ erg
sec$^{-1}$, respectively. These values take into account the
observational incompleteness for the Type II population in La Franca's
X-Ray sample ($\sim$1/3 of the sources would not observed). Table 1
shows that our results give an estimate of $7.0 \alt \gamma \alt 9.0$,
roughly coinciding with $\gamma$ for the lowest and middle luminosity
bins in La Franca's sample.

Unfortunately, the current statistics of the observed $N_H$
distribution does not allow us to test such luminosity dependency.
However, the number density of the most luminous sources is much
smaller than those of low and intermediate luminosity, with fractions
of 63\%, 30\%, and 7\% for the $10^{42-43}$, $10^{43-44}$ and
$10^{44-45}$ erg sec$^{-1}$ luminosity bins (integrating $L_X$ from
$10^{42}$ to $10^{45}$ erg sec$^{-1}$ in the Ueda's luminosity
function). Therefore, the general $N_H$ distribution is clearly
dominated for the less powerful (Seyfert like) AGN, for which the
derived torus properties might be a good representation.

\subsection{Column Densities Uncertainties}

Our analysis is based on the distribution of the determined column
densities. From this measurements, it is not possible to determine how
far from the central source the absorption is produced. This implies
that overdensities of gas located closer than the graphite sublimation
radius will not have the assumed {\it{Gas-to-Dust}} ratio. This
possibility is consistent with the extreme variability of the column
density observed in the Seyfert galaxy NGC\,1365 (Risaliti et
al. \cite{Risaliti05}). Also, assuming higher metallicities gives a
harder photoelectric absorption, increasing the estimation of the
number of Type I sources. Finally, changing the parameter $R_V$ from
$3.1$ to $5.0$ produces small differences in our results.

A spectroscopic analysis such as the one conducted by Tozzi et
al. (\cite{tozzi}) is essential to estimate the distribution of column
densities since particular spectral features for each source can be
present in the data. Broad band estimations, using $Hardness$ $Ratios$
are not precise enough and are highly affected by the template assumed
for the X-Ray source. These uncertainties affect the best-fit model
parameters, and the error bars for these estimations. We conclude that
a direct fit to the spectral features is determinant at the time to
describe the obscuring properties of the AGN population.

\subsection{Torus Model Results}

Recent IR observations have derived compact torus structures in nearby
Seyfert II galaxies (Jaffe et al. \cite{jaffe}; Prieto et
al. \cite{prieto}, \cite{prieto05}; Swain et al. \cite{swain}) in
agreement with the sizes predicted for model 1, while for models 3 and
4 this physical scale can be freely adjusted.

It has been argued that a constant torus density distribution does not
explain the reemission IR spectrum in AGN, but the inclusion of a
clumpy medium solves this problem. Models 1 to 3 can be discretized
using Equation \ref{nhss}, while adopting small optical depths per
cloud allows predictions of Type I sources at $\la 10^{21}$ cm$^{-2}$,
avoiding a large dichotomy between Type II and Type I objects (see
Model 4 for details).

Exponential angular dependencies of the parameter $\rho$ (see Equation
\ref{nhss}) allow smooth density decrements from the equatorial plane
to the poles, so that a wide range of $N_H$ can be obtained, implying
that Model 1 and 3 are the most favorable obscuring structures. On the
other hand, radial dependencies for the models ($\beta \neq 0$) can
shrink the effective torus radius, increasing the inner equatorial
density, more according with dynamical simulations of a thick clumpy
torus (see Beckert \& Duschl \cite{be_du}).

\subsection{Highly absorbed sources}

According to the X-Ray analysis of Seyfert II galaxies presented by
Maiolino et al. (\cite{Maiolino}), the number of Compton Thick sources
represents a large fraction of the AGN population.  In this work, the
assumption of a maximum column density at $N_H=10^{25}$ cm$^{-2}$
allows us to estimate this highly absorbed population from our
geometrical assumptions (see Figure \ref{geo_models}).  We roughly
predict intrinsic Compton Thick source fractions of 27\%, 3\%, 26\%
and 58\% for Model 1, 2, 3 and 4 respectively. Work by Tozzi et
al. (\cite{tozzi}) estimate a lower limit for the Compton Thick
fraction of $\sim 10\%$ in the CDF-S, while Beckmann et
al. (\cite{beckmann}) find 4 Compton Thick sources of a total of 40
AGN observed with $INTEGRAL$ in the 20-40 keV energy range at $z \sim
0$.  These values are somewhat smaller, but still consistent with
predictions from Models 1 and 3 given the uncertainties involved.

\section{Conclusions}
\begin{figure*}
  \centering
  \includegraphics[scale=0.8]{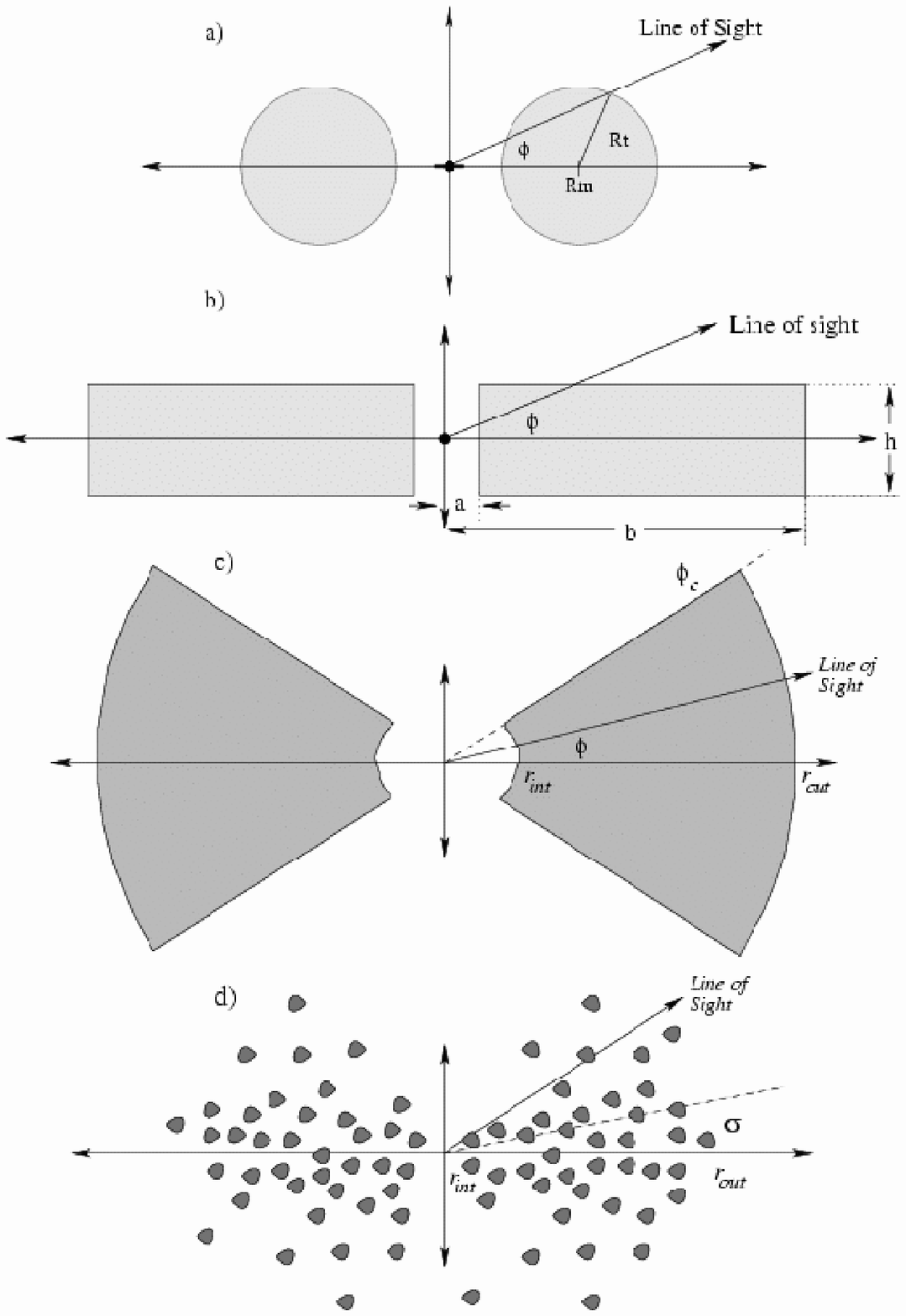}
  \caption
      {
	Geometrical matter density distributions assumed for the torus
	models. Figures a$)$, b$)$ c$)$ and d$)$ are based on the
	previous work by Treister et al. (\cite{Treister}), Pier \&
	Krolik (\cite{pi_kr92}), Granato \& Danese (\cite{gr_da}) and
	Nenkova et al. (\cite{Nenkova}), respectively.
      }
      \label{torus}
\end{figure*}

\begin{figure*}
  \centering
  \includegraphics[scale=0.44]{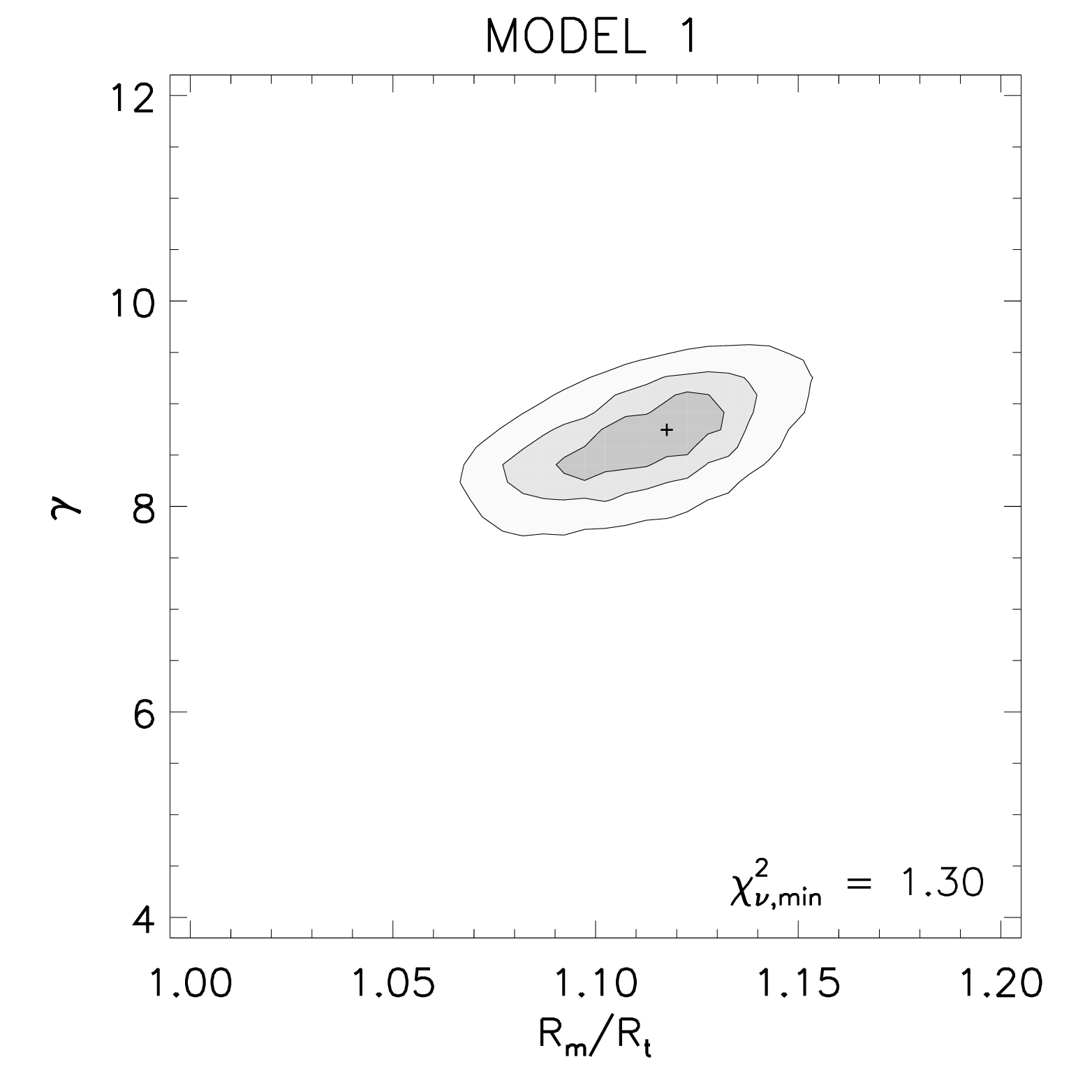}
  \includegraphics[scale=0.44]{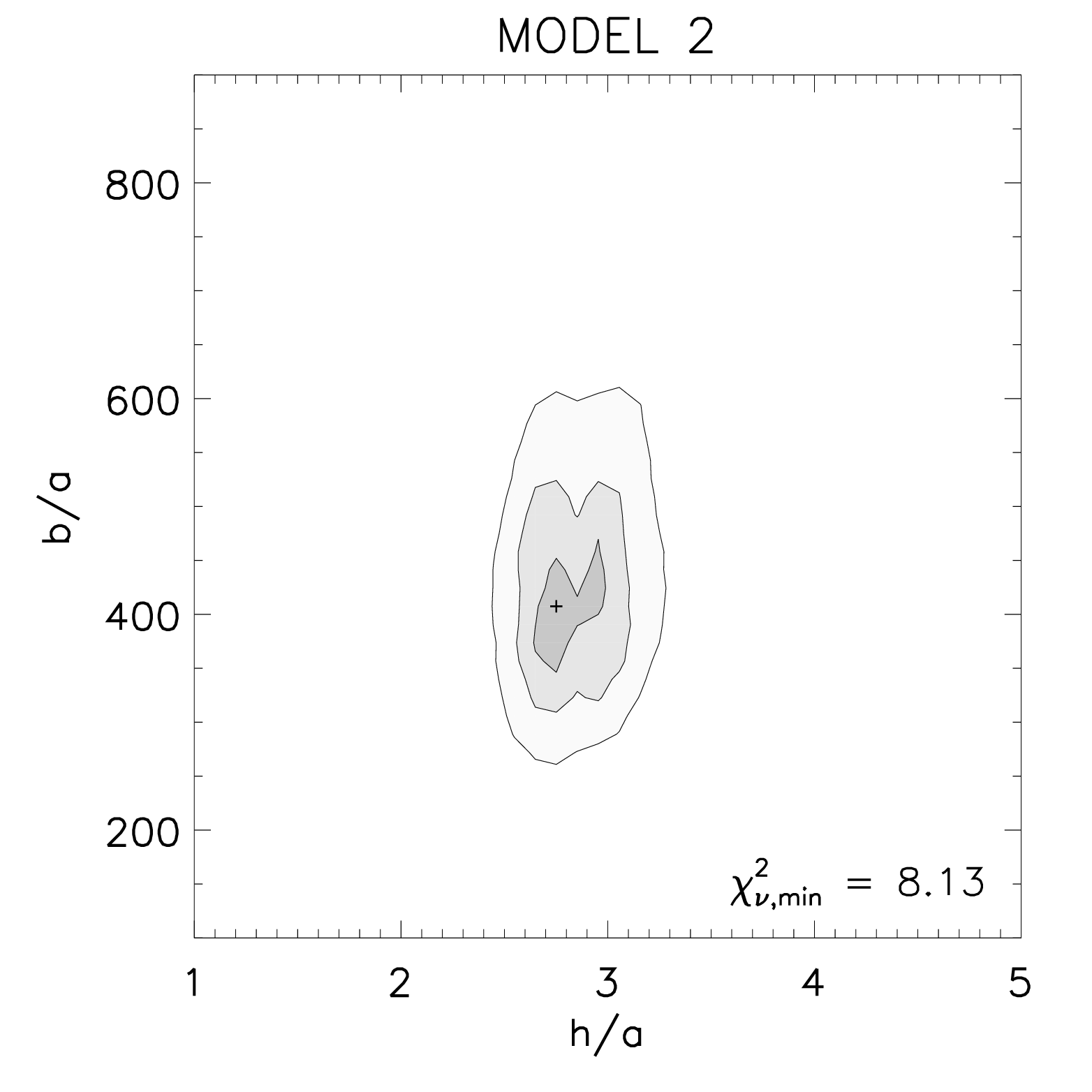}
  \includegraphics[scale=0.44]{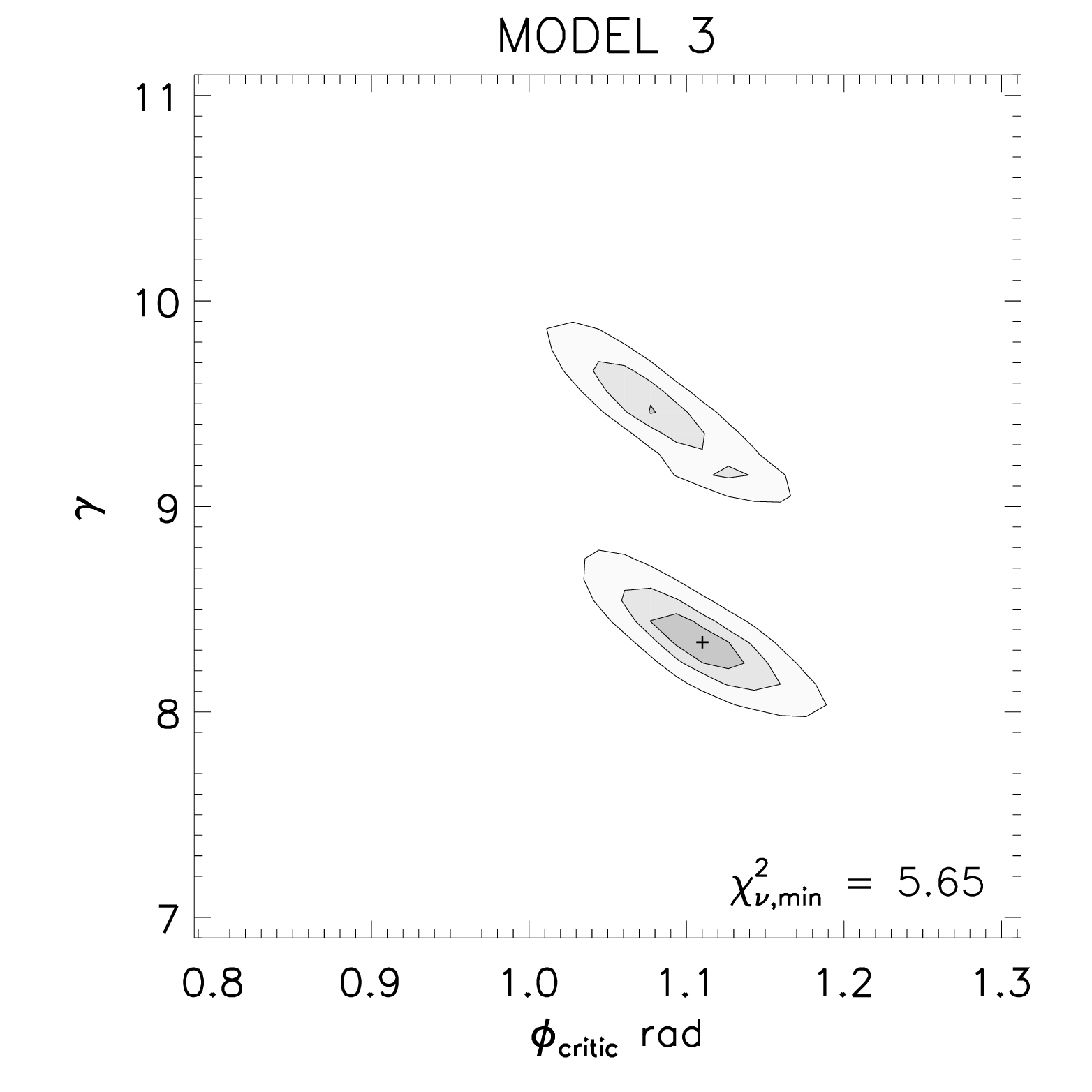}
  \includegraphics[scale=0.44]{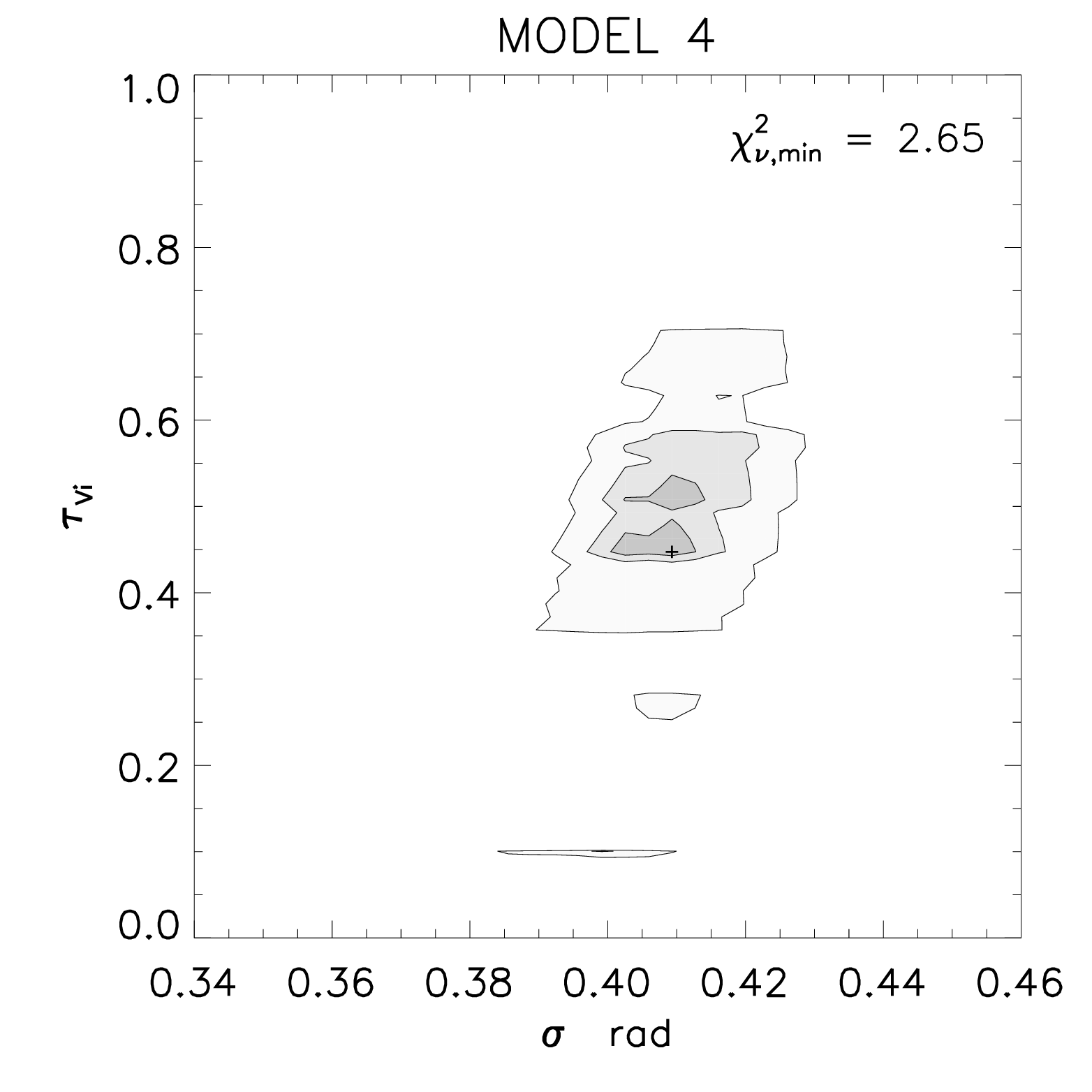}
  \caption
      { Best-fit parameters found for each model following Section
	\ref{torus_models}. Contours represent 1, 2 and 3 $\sigma$ 
	confidence limits. Note that for Model 1 $\beta=0.0$. 
      }
      \label{ks_models}
\end{figure*}

\begin{figure*}
  \centering
  \includegraphics[scale=0.34]{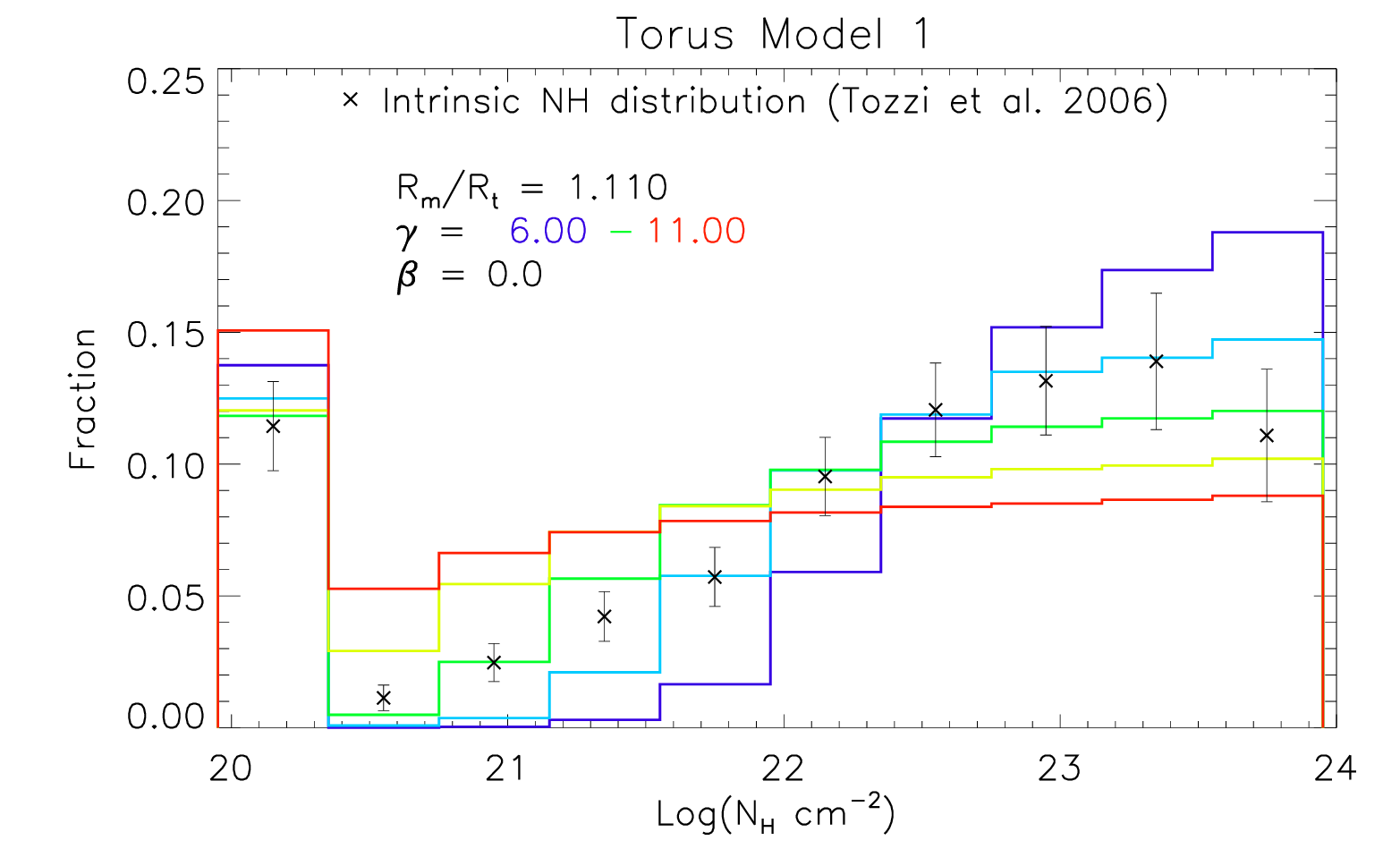}
  \includegraphics[scale=0.34]{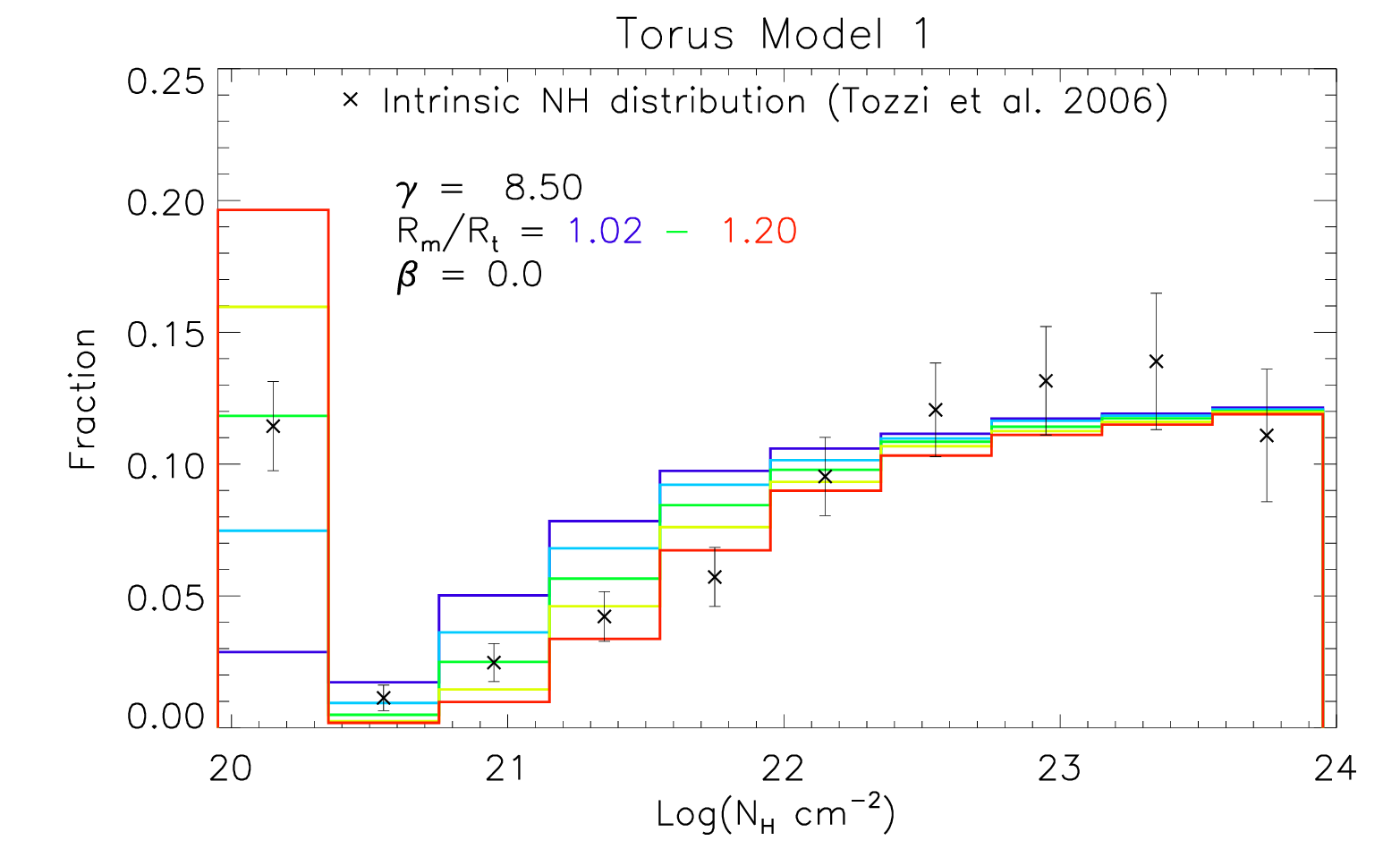}
  \includegraphics[scale=0.34]{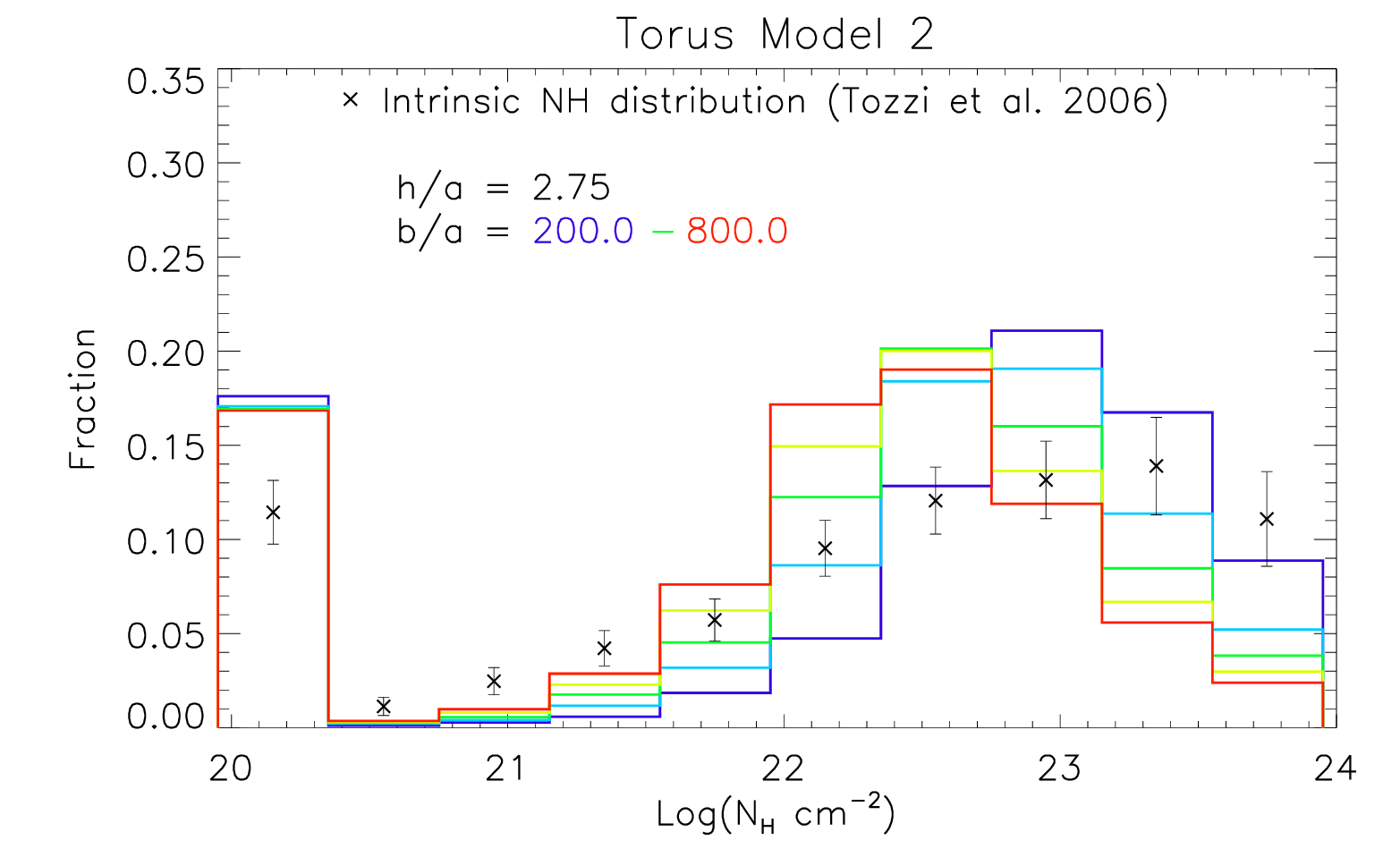}
  \includegraphics[scale=0.34]{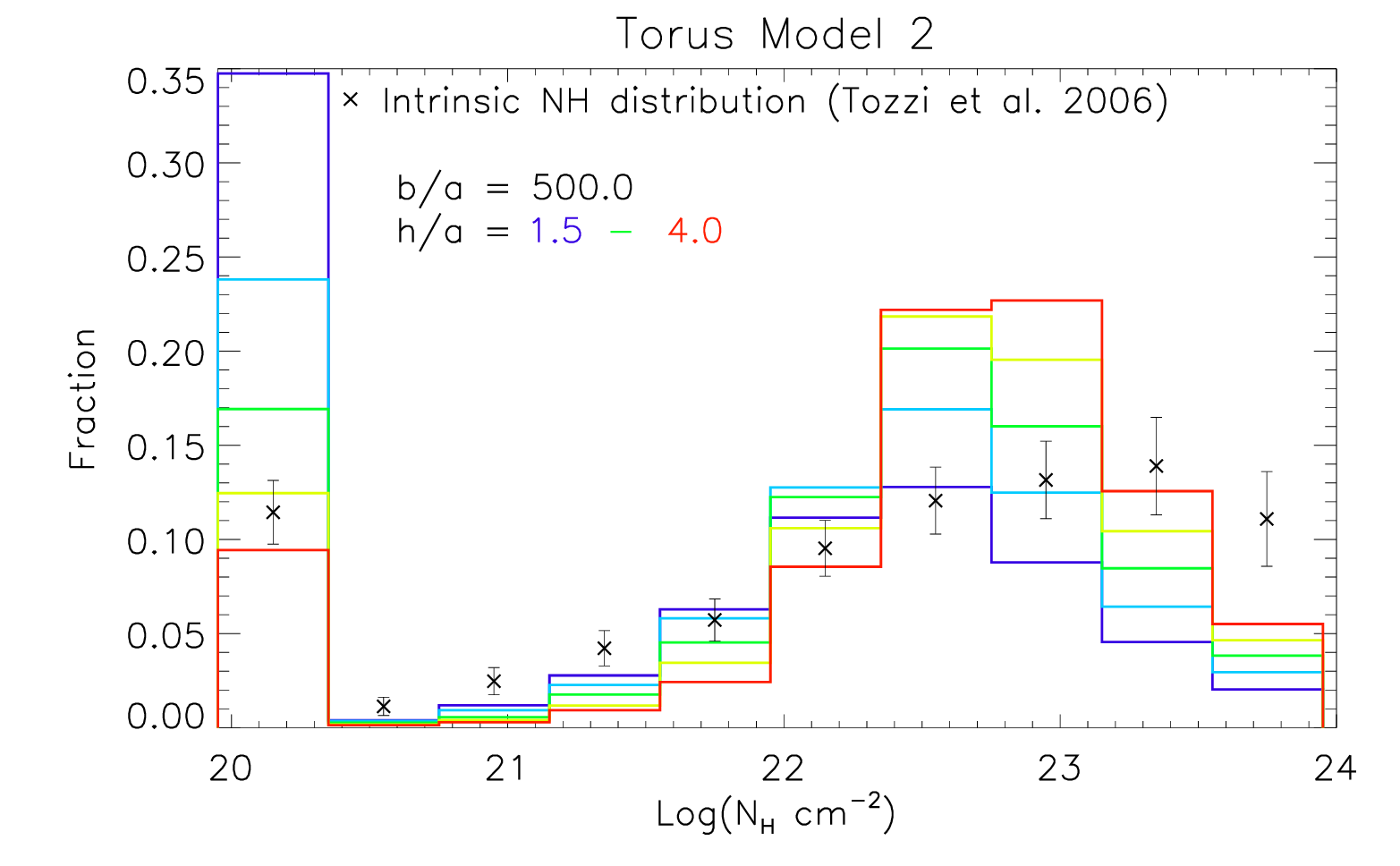}
  \includegraphics[scale=0.34]{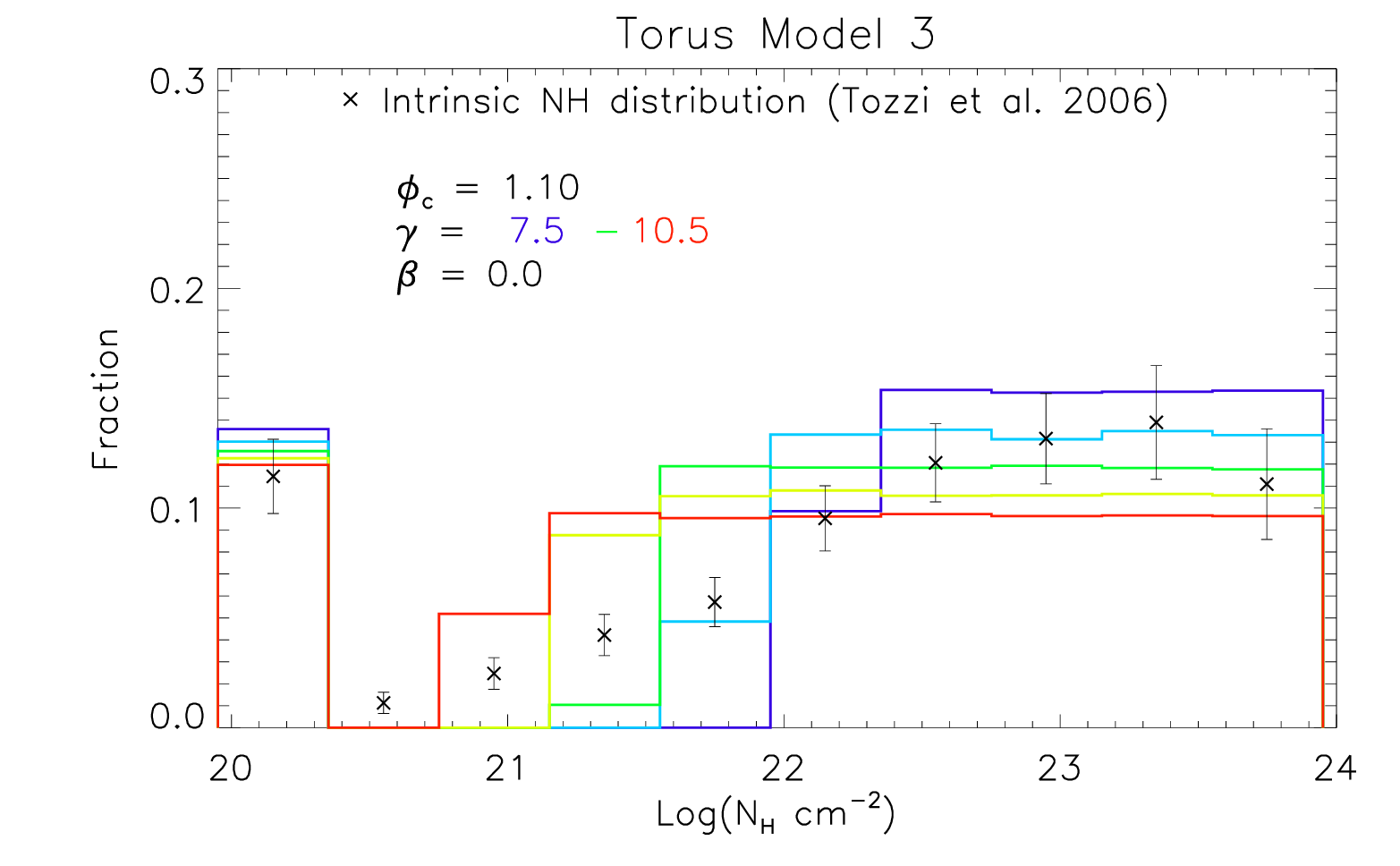}
  \includegraphics[scale=0.34]{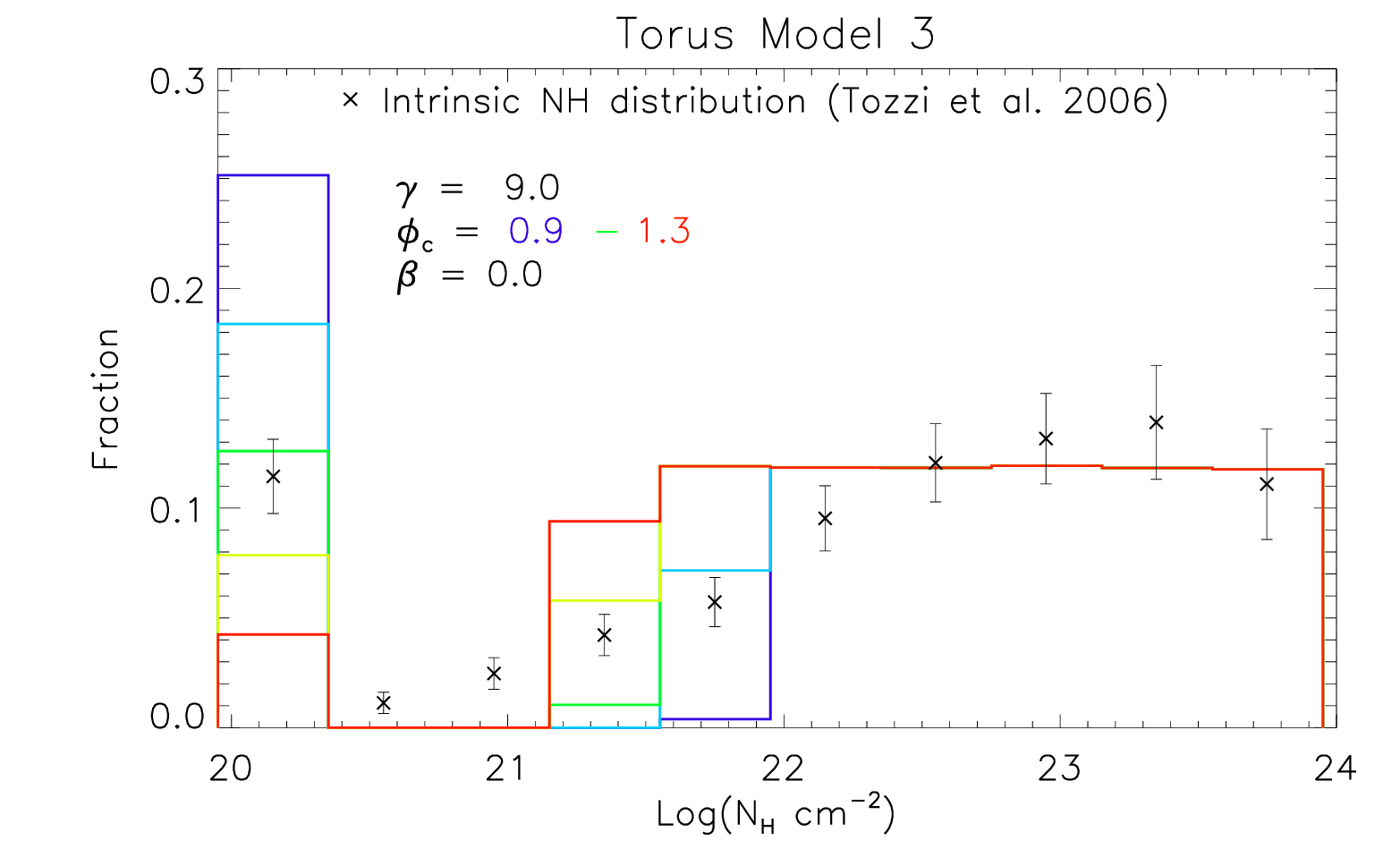}
  \includegraphics[scale=0.34]{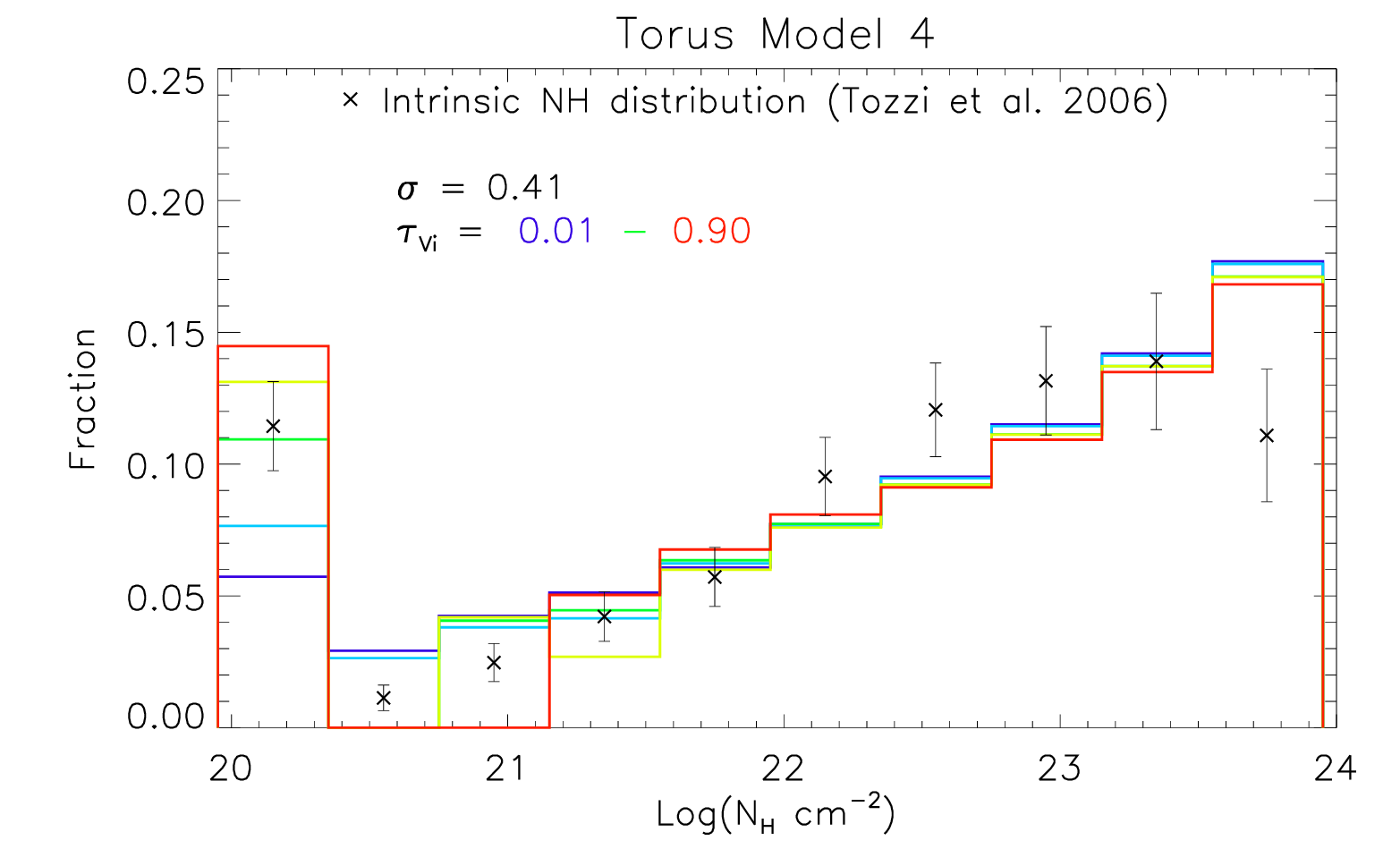}
  \includegraphics[scale=0.34]{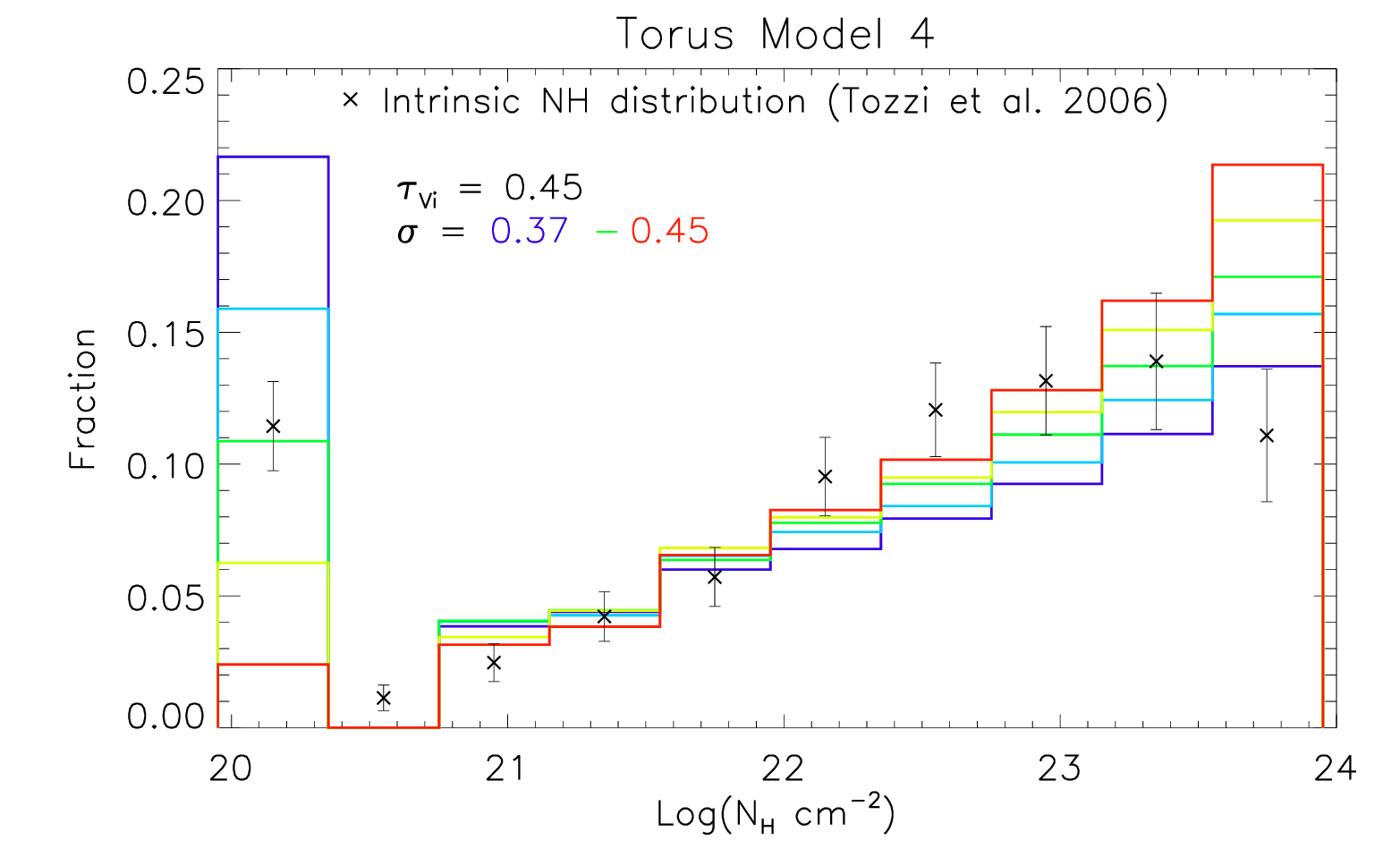}
  \caption
      { Comparison between the column densities of the observed
	$Chandra$ sources in the CDF-S field (crosses) and the best-fit
	results obtained from our modelling. For each model, two
	histograms are presented varying the range of interest of one
	parameter only.} 
      \label{best_models}
\end{figure*}

\begin{figure*}
  \centering
  \includegraphics[scale=0.44]{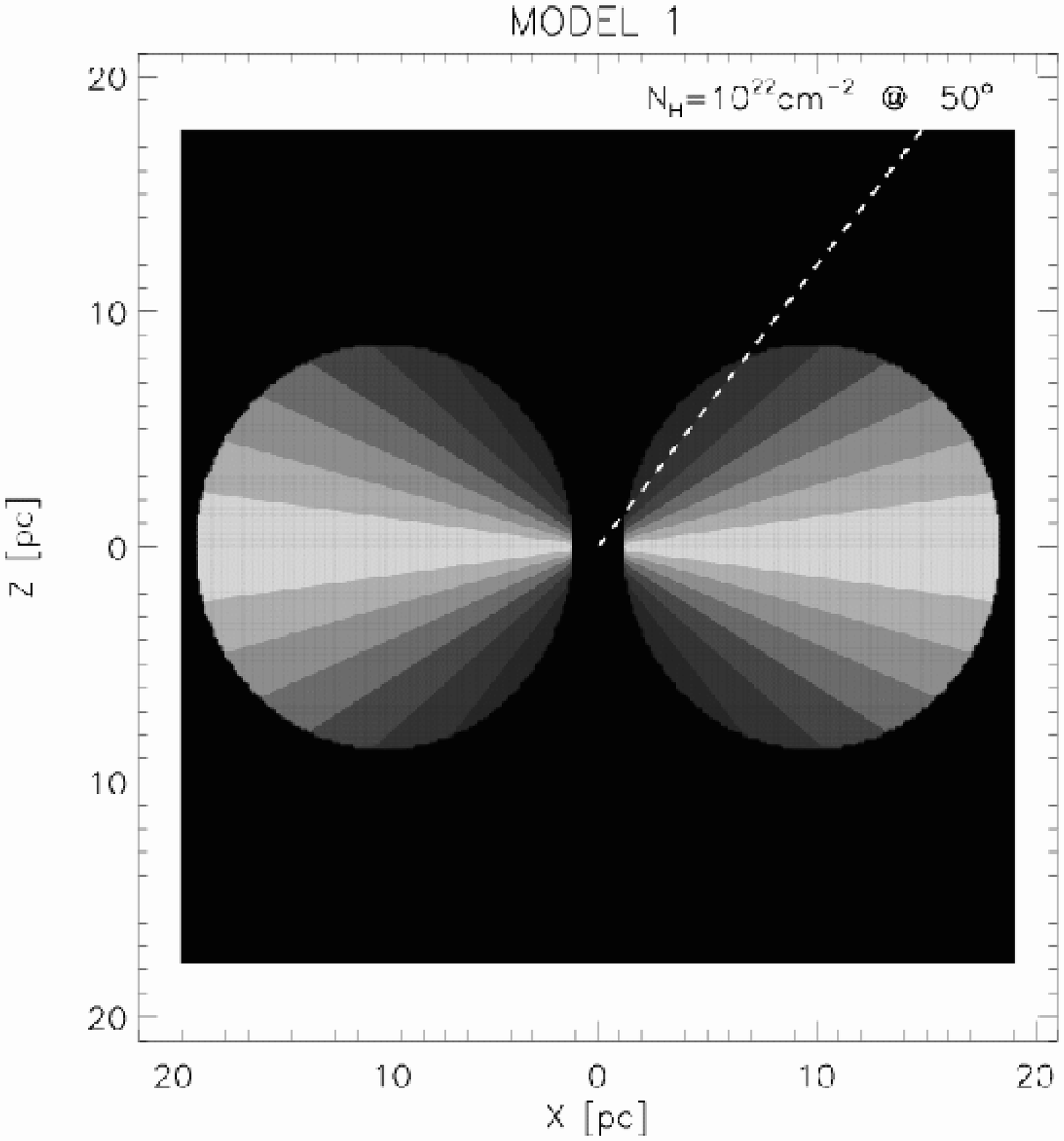}
  \includegraphics[scale=0.44]{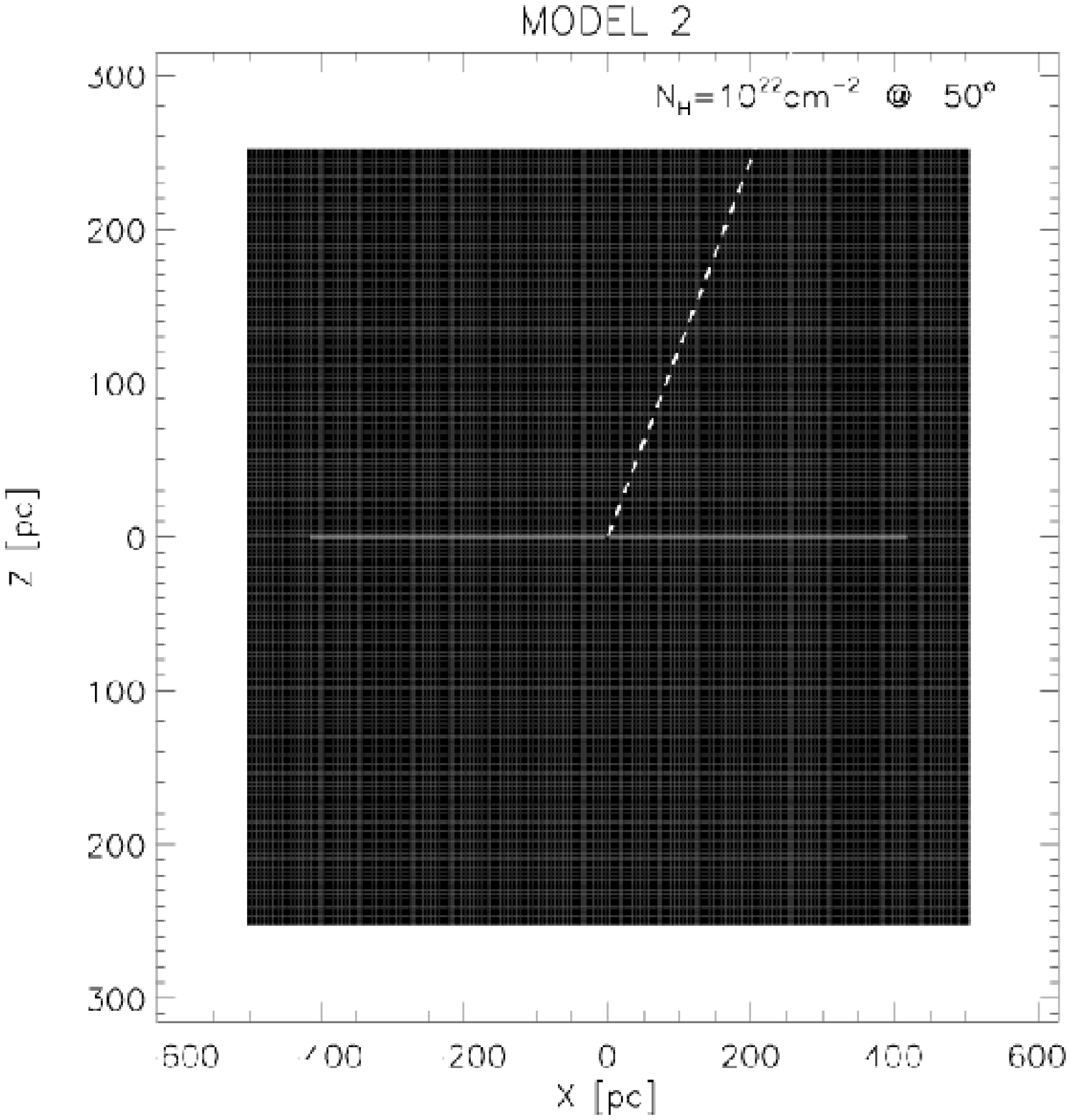}
  \includegraphics[scale=0.44]{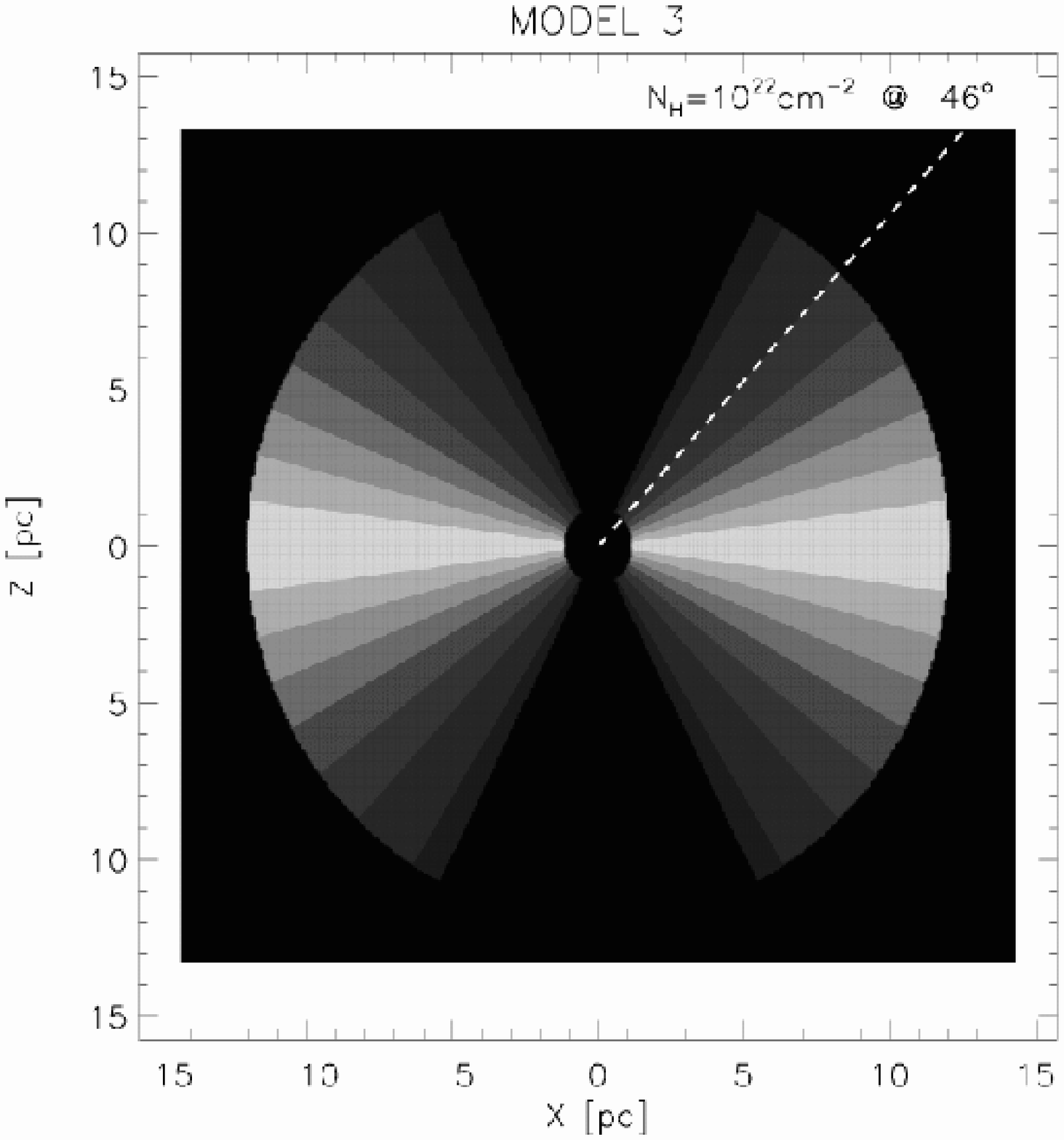}
  \includegraphics[scale=0.44]{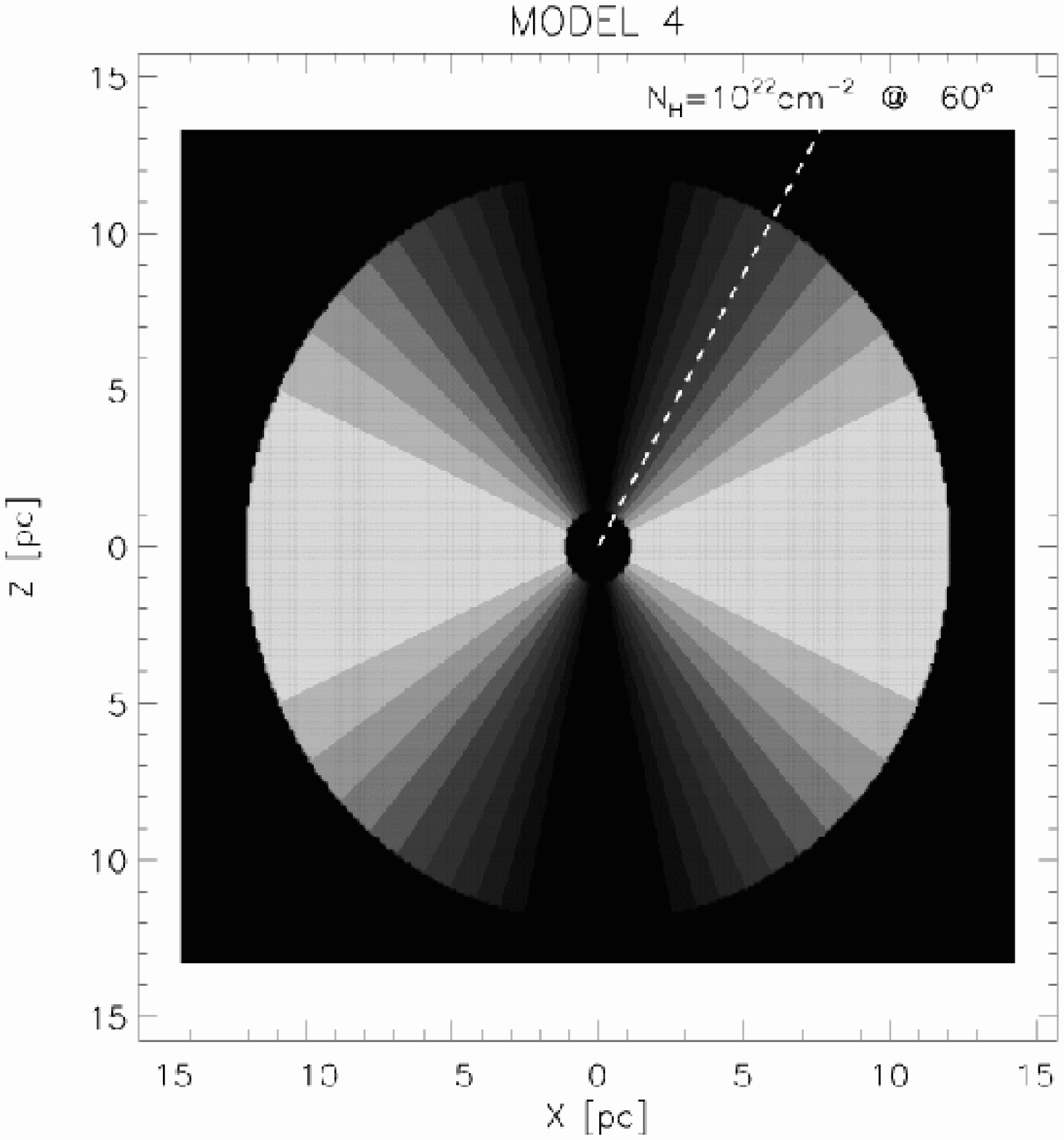}
  \caption
      {
	Geometrical matter density distribution using best fit parameters for
	each model. Model 1: $R_m/R_t=1.14$, $\gamma=9.0$ and
	$\beta=0.0$; Model 2: $h/a=3.0$ and $b/a=350$; Model 3:
	$\phi_c=1.1$ and $\gamma=9.5$; Model 4: $\sigma=0.4$ and
	$\tau_{Vi}<<1$ (continuous case). The inner
	radius is set at 1.2 pc  for all models, and an outer radius at
	12 pc for models 3 and 4 (but adjustable for more compact
	structures). Different shades mean $\Delta\log(N_H)=0.5$, starting
	from $10^{25}$ cm$^{-2}$ at the equatorial plane. The dashed
	line represents the line of sight that divides the Type I and
	the Type II sources. Note that Model 2 is shown at a different 
	scale. 
      }
      \label{geo_models}
\end{figure*}

\begin{enumerate}

\item
  Using the observed distribution of $N_H$ columns for AGNs in the
  CDF-S, we have constrained the geometry of the obscuring region
  present in AGNs. Four different torus geometries, based on
  previously proposed models (see Figure \ref{torus}), have been
  explored. No luminosity or redshift dependency has been included in
  the torus modelling.
  
\item
  Results from our fit to the $N_H$ distribution are presented in
  Table \ref{table1} which shows the model parameter best-fit
  values. These results can be used to estimate physical properties of
  the torus, such as the inner and outer radius, and the density at
  the equatorial plane, as a function of the AGN luminosity.

\item 
  Detailed analysis shows that a matter density profile ($\rho(\phi)$)
  with an exponential dependency in $\phi$, such as in Model 1 and 3,
  gives the best representation of the wide range of observed column
  densities (from $\sim 10^{20}$ cm$^{-2}$ to $\sim 10^{25}$
  cm$^{-2}$). A constant density distribution (Model 2) requires very
  extended structures to recover the wide observed $N_H$ range (since
  $N_H\propto \Delta s$), not according with actual observations of
  local Seyfert galaxies where pc-scale torus-like structures have
  been detected. It also underestimates the fraction of Type II sources. A
  Gaussian angular dependency (Model 4), as in the 
  clumpy model proposed by Nenkova et al. (\cite{Nenkova}), results in
  a substantial overestimation of highly obscured sources. We also
  find that the optical depth for individual clumps, $\tau_{Vi}$, has
  to be $\sim 60$ times smaller than the value adopted by Nenkova to
  compute the torus IR reemission. We note that the clumpy torus
  treatment could also be applied to other model distributions using
  Equation \ref{nhss} and a suitable value for $\tau_{Vi}$.  Models 3
  and 4 suffer of multiple $\chi^2$ minima caused by the dichotomy
  between Type I and Type II sources (given by the critical value of
  $\phi$ and the presence of at least 1 clump along the line of sight,
  respectively).

\item
  We therefore conclude that Model 1, a classical `donut shape', is
  the best parameterization for the obscuring region around AGN. Since
  no luminosity dependency has been included in our modelling, the
  results can be regarded as a luminosity averaged parameterization of
  the torus, which is largely dominated by low and intermediate luminosity
  sources.
 
\end{enumerate}

\begin{acknowledgements}
  This paper was supported by Proyecto MECESUP {\it{``Expansi\'on del
      Universo Astron\'omico en Chile''}} and by GEMINI-PPARC
      research studentship. Thanks to E. Treister,
  D. Gruber, R. Wilman, D. Alexander, P. Best and R. Ivison for their
  data and comments for this work. P. Lira acknowledges support from
  Fondecyt project N$^{\texto{o}}$ 1040719. {\it{E.Ibar, agradece
      a su familia, amigos y como no tambi\'en al SAJ.}}

\end{acknowledgements}

\newpage

\newpage
 
\newpage

\newpage

\end{document}